\documentclass[twoside,twocolumn,9pt]{article}
\usepackage{extsizes}
\usepackage[super,sort&compress,comma]{natbib} 
\usepackage[version=3]{mhchem}
\usepackage[left=1.5cm, right=1.5cm, top=1.785cm, bottom=2.0cm]{geometry}
\usepackage{balance}
\usepackage{mathptmx}
\usepackage{sectsty}
\usepackage{graphicx} 
\usepackage{lastpage}
\usepackage[format=plain,justification=justified,singlelinecheck=false,font={stretch=1.125,small,sf},labelfont=bf,labelsep=space]{caption}
\usepackage{float}
\usepackage{fancyhdr}
\usepackage{fnpos}
\usepackage[english]{babel}
\usepackage{array}
\usepackage{droidsans}
\usepackage{charter}
\usepackage[T1]{fontenc}
\usepackage[usenames,dvipsnames]{xcolor}
\usepackage{setspace}
\usepackage[compact]{titlesec}
\usepackage{hyperref}

\usepackage{epstopdf}

\definecolor{cream}{RGB}{222,217,201}

\begin{document}

\pagestyle{fancy}
\thispagestyle{plain}
\fancypagestyle{plain}{
\renewcommand{\headrulewidth}{0pt}
}

\makeFNbottom
\makeatletter
\renewcommand\LARGE{\@setfontsize\LARGE{15pt}{17}}
\renewcommand\Large{\@setfontsize\Large{12pt}{14}}
\renewcommand\large{\@setfontsize\large{10pt}{12}}
\renewcommand\footnotesize{\@setfontsize\footnotesize{7pt}{10}}
\makeatother

\renewcommand{\thefootnote}{\fnsymbol{footnote}}
\renewcommand\footnoterule{\vspace*{1pt}%
\color{cream}\hrule width 3.5in height 0.4pt \color{black}\vspace*{5pt}} 
\setcounter{secnumdepth}{5}

\makeatletter 
\renewcommand\@biblabel[1]{#1}            
\renewcommand\@makefntext[1]%
{\noindent\makebox[0pt][r]{\@thefnmark\,}#1}
\makeatother 
\renewcommand{\figurename}{\small{Fig.}~}
\sectionfont{\sffamily\Large}
\subsectionfont{\normalsize}
\subsubsectionfont{\bf}
\setstretch{1.125} 
\setlength{\skip\footins}{0.8cm}
\setlength{\footnotesep}{0.25cm}
\setlength{\jot}{10pt}
\titlespacing*{\section}{0pt}{4pt}{4pt}
\titlespacing*{\subsection}{0pt}{15pt}{1pt}

\fancyfoot{}
\fancyfoot[LO,RE]{\vspace{-7.1pt}\includegraphics[height=9pt]{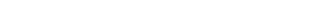}}
\fancyfoot[CO]{\vspace{-7.1pt}\hspace{13.2cm}\includegraphics{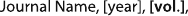}}
\fancyfoot[CE]{\vspace{-7.2pt}\hspace{-14.2cm}\includegraphics{head_foot/RF}}
\fancyfoot[RO]{\footnotesize{\sffamily{1--\pageref{LastPage} ~\textbar  \hspace{2pt}\thepage}}}
\fancyfoot[LE]{\footnotesize{\sffamily{\thepage~\textbar\hspace{3.45cm} 1--\pageref{LastPage}}}}
\fancyhead{}
\renewcommand{\headrulewidth}{0pt} 
\renewcommand{\footrulewidth}{0pt}
\setlength{\arrayrulewidth}{1pt}
\setlength{\columnsep}{6.5mm}
\setlength\bibsep{1pt}

\makeatletter 
\newlength{\figrulesep} 
\setlength{\figrulesep}{0.5\textfloatsep} 

\newcommand{\topfigrule}{\vspace*{-1pt}%
\noindent{\color{cream}\rule[-\figrulesep]{\columnwidth}{1.5pt}} }

\newcommand{\botfigrule}{\vspace*{-2pt}%
\noindent{\color{cream}\rule[\figrulesep]{\columnwidth}{1.5pt}} }

\newcommand{\dblfigrule}{\vspace*{-1pt}%
\noindent{\color{cream}\rule[-\figrulesep]{\textwidth}{1.5pt}} }

\makeatother

\twocolumn[
  \begin{@twocolumnfalse}
{\includegraphics[height=30pt]{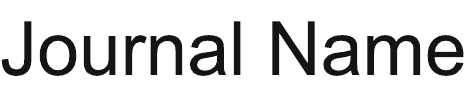}\hfill\raisebox{0pt}[0pt][0pt]{\includegraphics[height=55pt]{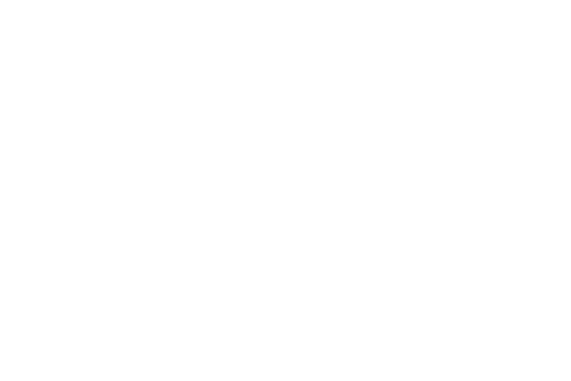}}\\[1ex]
\includegraphics[width=18.5cm]{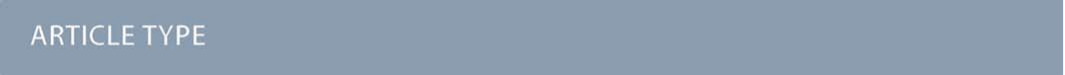}}\par
\vspace{1em}
\sffamily
\begin{tabular}{m{4.5cm} p{13.5cm} }

\includegraphics{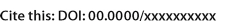} & \noindent\LARGE{\textbf{Assisting the Grading of a Handwritten\newline General Chemistry Exam with Artificial Intelligence}} \\
\vspace{0.3cm} & \vspace{0.3cm} \\

 & \noindent\large{Jan Cvengros,$^{\ast}$\textit{$^{a}$} and Gerd Kortemeyer$^{\ast}$\textit{$^{b\dag}$}} \\

\includegraphics{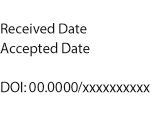} & \noindent\normalsize{We explore the effectiveness and reliability of an artificial intelligence (AI)-based grading system for a handwritten general chemistry exam, comparing AI-assigned scores to human grading across various types of questions. Exam pages and grading rubrics were uploaded as images to account for chemical reaction equations, short and long open-ended answers, numerical and symbolic answer derivations, drawing, and sketching in pencil-and-paper format. Using linear regression analyses and psychometric evaluations, the investigation reveals high agreement between AI and human graders for textual and chemical reaction questions, while highlighting lower reliability for numerical and graphical tasks. The findings emphasize the necessity for human oversight to ensure grading accuracy, based on selective filtering. The results indicate promising applications for AI in routine assessment tasks, though careful consideration must be given to student perceptions of fairness and trust in integrating AI-based grading into educational practice.}\end{tabular}

 \end{@twocolumnfalse} \vspace{0.6cm}

  ]

\renewcommand*\rmdefault{bch}\normalfont\upshape
\rmfamily
\section*{}
\vspace{-1cm}


\footnotetext{\textit{$^{a}$~Department of Chemistry and Applied Biosciences, ETH Zurich, HCI H 101, Vladimir-Prelog-Weg 1-5/10, 8093 Z{\"u}rich, Switzerland. E-mail: cvengros@chem.ethz.ch}}
\footnotetext{\textit{$^{b}$~Rectorate and ETH AI Center, ETH Zurich, R{\"a}mistrasse 101, 8092 Z{\"u}rich, Switzerland. E-mail: kgerd@ethz.ch}}
\footnotetext{\textit{$^{\dag}$}~also at Michigan State University, East Lansing, Michigan 48824, USA}


\section{Introduction}
\subsection{Open-ended responses}
In many chemistry courses, assessment technologies remain anchored in closed-answer formats: multiple choice, short-answer numerical, or very short narrative responses~\cite{mullen2012short}. To probe argumentation and reasoning, instructors often must recast tasks into these closed formats to accommodate the technology~\cite{stowe2019assessment}. While there have been sustained efforts to embed higher-order thinking into introductory chemistry exams~\cite{zoller2002algorithmic}, assessment techniques such as oral examinations (one robust avenue for open-ended reasoning) and open-ended written responses are rarely scalable in large-enrollment settings~\cite{gardner2023challenges}. Yet open-ended problem solving is central to STEM assessment, offering rich visibility into conceptual understanding~\cite{alsalmani23} and reasoning processes~\cite{yik2021development,martin2023machine}. Compounding these tensions, the pandemic era amplified concerns about academic integrity in remote settings~\cite{clark2020testing}.

Outside exam settings, frequent formative assessment further enhances learning through timely feedback~\cite{laverty12b,offerdahl2019formative}, but it is often underused because of grading workload. What students focus on is oftentimes driven by what is assessed and how much credit is assigned to the assessment, where ungraded exercises may be neglected~\cite{gomez2025student}. Traditional automated systems, generally limited to closed-format items (e.g., multiple choice or numeric responses), can yield strong psychometric reliability~\cite{pawl2013} but struggle to capture higher-order skills and can invite guessing~\cite{towns2014guide}. Longer answers, including creative exercises~\cite{grieger2025utility}, mathematics-intensive exercises~\cite{ye2025just}, drawing tasks~\cite{baade2025seeing,stammes2025drawing}, and minute-papers~\cite{pentucci2025developing}, can foster formative self-reflection and help students in discovering knowledge structures and possible misconceptions.

\subsection{Automated grading}
Against this backdrop, the rapid evolution of artificial intelligence, particularly large language models (LLMs) such as generative Pre-trained Transformers (GPT), is beginning to expand assessment options. Beyond scaling existing practices~\cite{lawrie2023establishing}, recent work indicates that LLMs can evaluate complex, open-ended student responses at scale~\cite{meyer2023chatgpt,kasneci2023chatgpt,wilson22,wan24,kortemeyer24aigrading,kortemeyer2024grading}. By reducing grading burden, these systems may make more frequent formative and low-stakes summative assessments feasible without overwhelming instructional resources~\cite{Floden2025,Fernandez2024}. In short, while current technologies have historically nudged assessments toward closed formats, advances in AI offer credible pathways to recover and systematically assess the richer reasoning and explanatory work that open-ended chemistry problems demand.

Despite these potential benefits, integrating AI into grading workflows introduces critical considerations regarding grading reliability and transparency. AI-generated scores can occasionally exhibit undue confidence in ambiguous cases or misinterpret nuanced responses~\cite{ding2023students,dahlkemper23,li2023wrong}. Regulatory frameworks, such as the European Union's designation of AI in educational assessments as high-risk, mandate human oversight to address these reliability concerns \cite{euactannex,euactobl,de2020case}. Consequently, a human-in-the-loop grading approach is essential, ensuring that complex or ambiguous student responses receive necessary human attention~\cite{alsalmani23a,li2023learning}.

Grading provides valuable pedagogical benefits for teaching assistants (TAs), which may be an argument against using these systems. After all, by evaluating student work --- particularly ambiguous or challenging cases --- TAs deepen their own content knowledge and pedagogical understanding~\cite{cho2010learning}. AI-assisted grading systems can beneficially direct TA attention precisely toward these more instructive cases, filtering out routine grading tasks that offer limited educational value.

This study investigates AI-assisted grading within a high-stakes, handwritten chemistry exam context. We employ psychometric methods, including Item Response Theory (IRT)~\cite{rasch}, to evaluate the reliability of AI-generated scores, developing confidence thresholds that identify when human intervention is required. While our immediate focus is on high-stakes assessments, the methodologies presented are equally applicable to lower-stakes and purely formative scenarios, supporting the practical expansion of frequent assessments. Ultimately, this approach aims to balance efficiency with educational quality, promoting scalable, fair, and pedagogically rich assessment practices.

\subsection{Theoretical framework}
We ground our approach in test theory and generalizability theory, treating AI-assigned points as fallible observations with multiple error sources (items, raters/runs, tasks) that can be estimated and, where appropriate, reduced through aggregation or design \cite{brennan1992generalizability}. For the interpretation and use of AI-assisted scores, we adopt Kane's argument-based validity perspective, which requires that claims about score meaning (and decisions such as pass/fail cut scores) be supported by explicit warrants and evidence~\cite{kane2013validating}. Our human-in-the-loop setup follows selective-prediction/deferral frameworks: the model grades confidently when reliable and defers uncertain cases to humans, analogous to the efforts in automated short-answer grading~\cite{li2023learning}.

\subsection{Related efforts in AI-assisted grading}
Within chemistry, early pilots have begun to use generative AI for automatic scoring of open-ended conceptual responses~\cite{yamtinah2024_icce,sonkar2024automated}, and complementary domain-specific approaches have automated evaluation of reaction answers within learning platforms \cite{plyer2024_softgrading}. Across STEM, controlled studies report that large language models (LLMs) such as GPT-3.5/4 can score open-ended or short-answer exam responses with high agreement to human raters~\cite{impey2025using,chen2025grading} and improve further with fine-tuning and chain-of-thought prompting~\cite{lee2024_patterns,latif2024_patterns}. In higher-education settings, multi-course deployments have graded thousands of free-text answers (in multiple languages) using LLMs guided by instructor rubrics~\cite{grevisse2024_bmcmeded,kortemeyer2025artificial,kortemeyer2024grading,kortemeyer2025assessing}. In science education, human-in-the-loop workflows combine LLM scoring with explanations and selective review to increase trustworthiness~\cite{cohn2024_cot}. Benchmarking work aggregates diverse short-answer datasets and finds LLM-based graders outperform earlier baselines~\cite{kortemeyer2023performance}, yet still benefit from calibration and rubric alignment~\cite{asag2024}. Recent educational-data-mining studies add retrieval or multi-agent designs to stabilize grading on science short answers \cite{chu2025_edm_graderag}, and multimodal pipelines now target scanned, diagram-rich or handwritten responses in science assessments \cite{zhu2025_ijcai}. Together, these strands indicate a rapid shift toward AI-assisted scoring of open-ended work in chemistry and related STEM domains, while reinforcing the need for transparent reliability checks and human oversight.

\section{Setting}
\subsection{Course}
Our study was carried out at ETH Zurich (German: Eidgen{\"os}sische Technische Hochschule Z{\"u}rich; English: Federal Institute of Technology Zurich), a public university located in Zurich, Switzerland. Founded in 1854, it specializes in science, technology, engineering, and mathematics. The course setting was a large-enrollment, general chemistry course for non-majors, in particular from biology and pharmaceutical sciences, as well as the health sciences and technology program. The course was taught and assessed in German, which is the default language for undergraduate courses at ETH Zurich.

\subsection{Exam}
\begin{figure*}
\centering
  \includegraphics[width=0.49\textwidth]{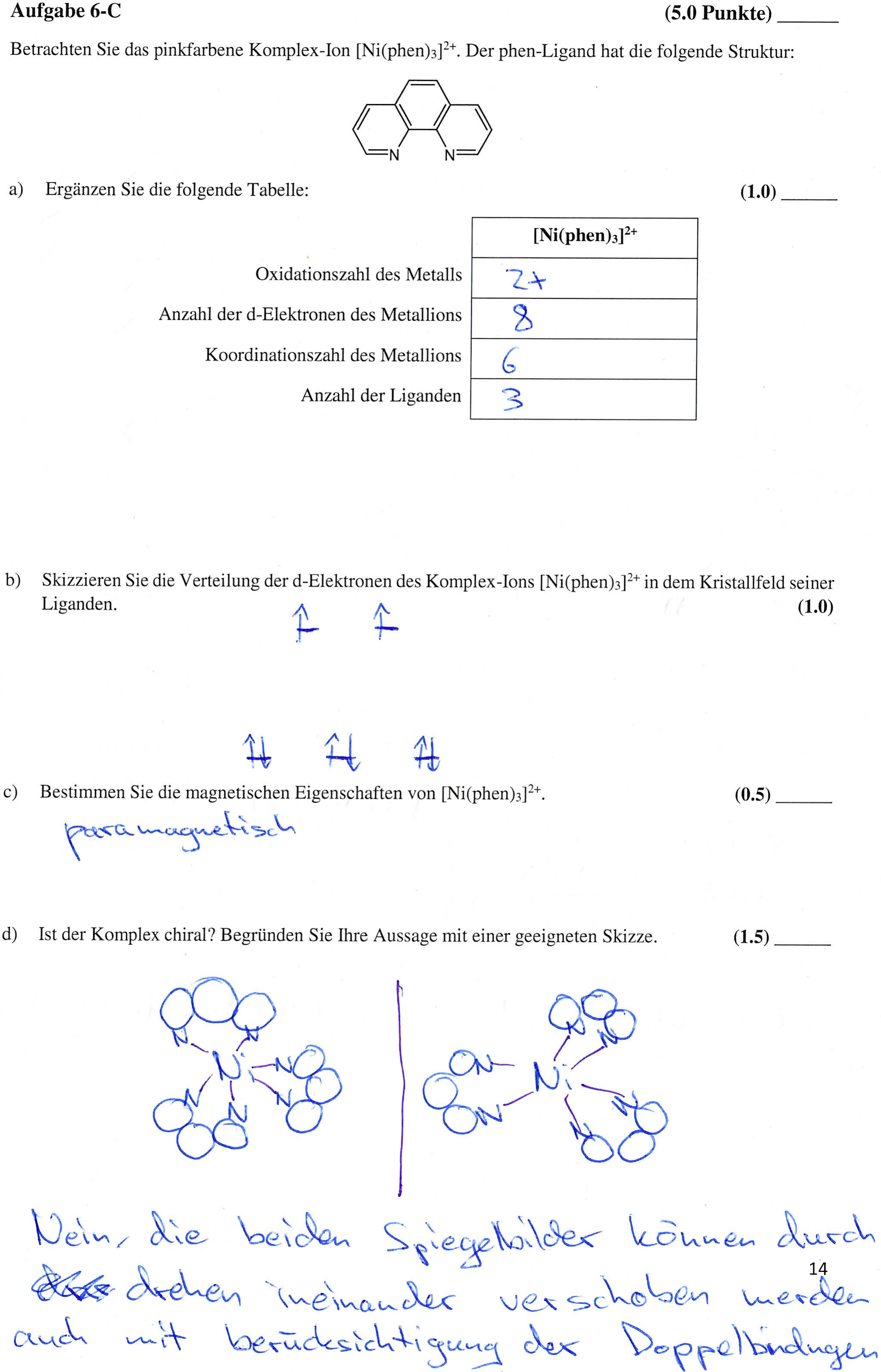}
  \includegraphics[width=0.49\textwidth]{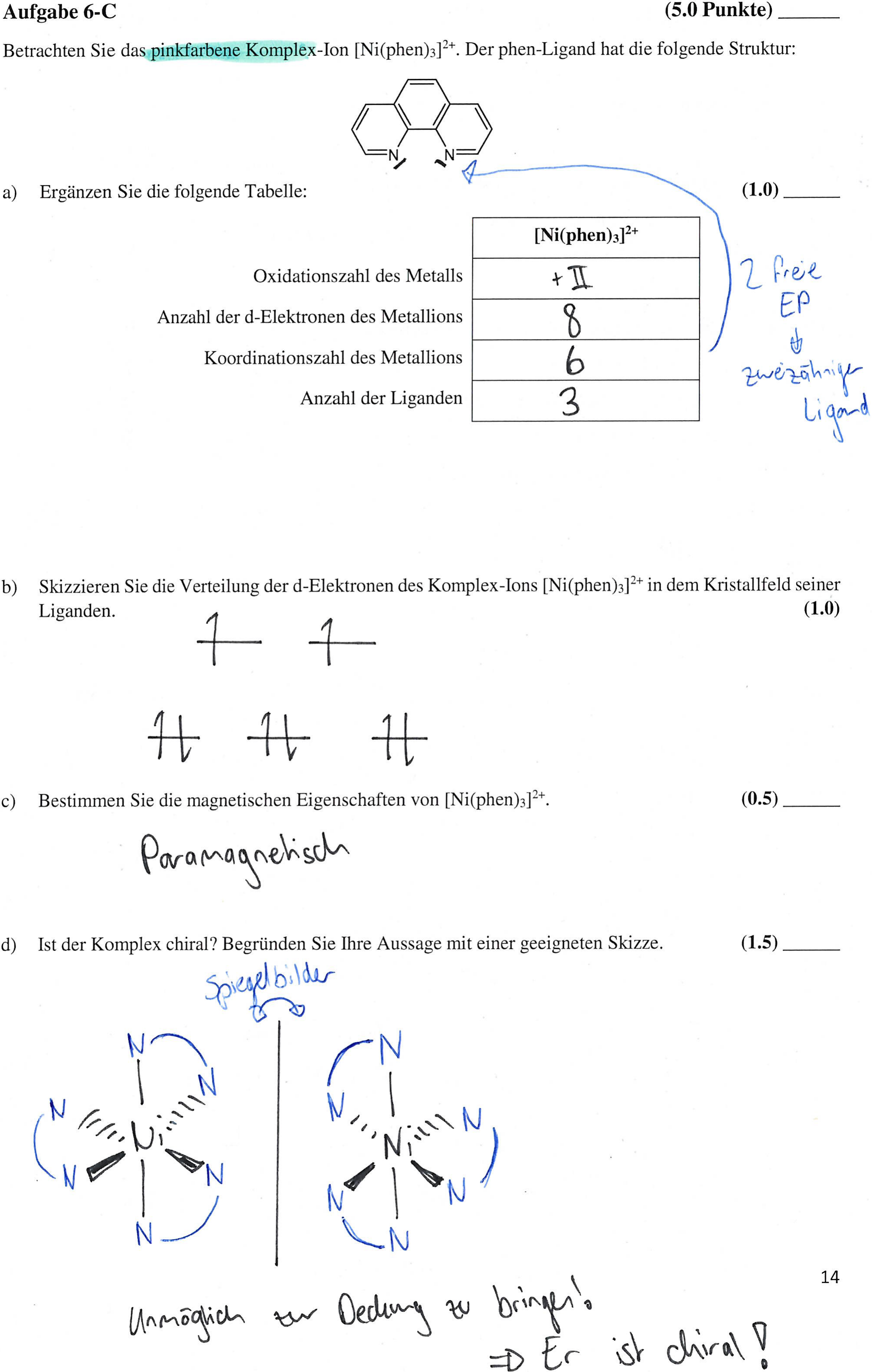}
  \caption{\ Examples of student work. These would be uploaded as images to the model. The exam was given in German, the translation of the problems is: Consider the pink-colored complex ion $[\mbox{Ni}(\mbox{phen})_3]^{2+}$. The phen-ligand has the following structure: \ldots
  Complete the following table: oxidation number of the metal, number of d-electrons of the metal ion, coordination number of the metal ion, number of ligands. Sketch the distribution of the d-electrons of the complex ion $[\mbox{Ni}(\mbox{phen})_3]^{2+}$ in the crystal field of its ligands. Determine the magnetic properties of  $[\mbox{Ni}(\mbox{phen})_3]^{2+}$. Is the complex chiral? Provide the reasoning of your answer with an appropriate sketch. The student work on the right labelled the initial image of the structure, ``2 free EP $\to$ bidentate ligand.'' The student's reasoning in the left panel is, ``no, the two mirror images can be superimposed by rotation, even when the double bonds are taken into account;'' the reasoning on the right is, after labelling the two sketches ``mirror images,'' ``impossible to superimpose! $\Rightarrow$ it is chiral!''}
  \label{fig:examples}
\end{figure*}

\begin{figure}
\centering

  \includegraphics[width=0.5\textwidth]{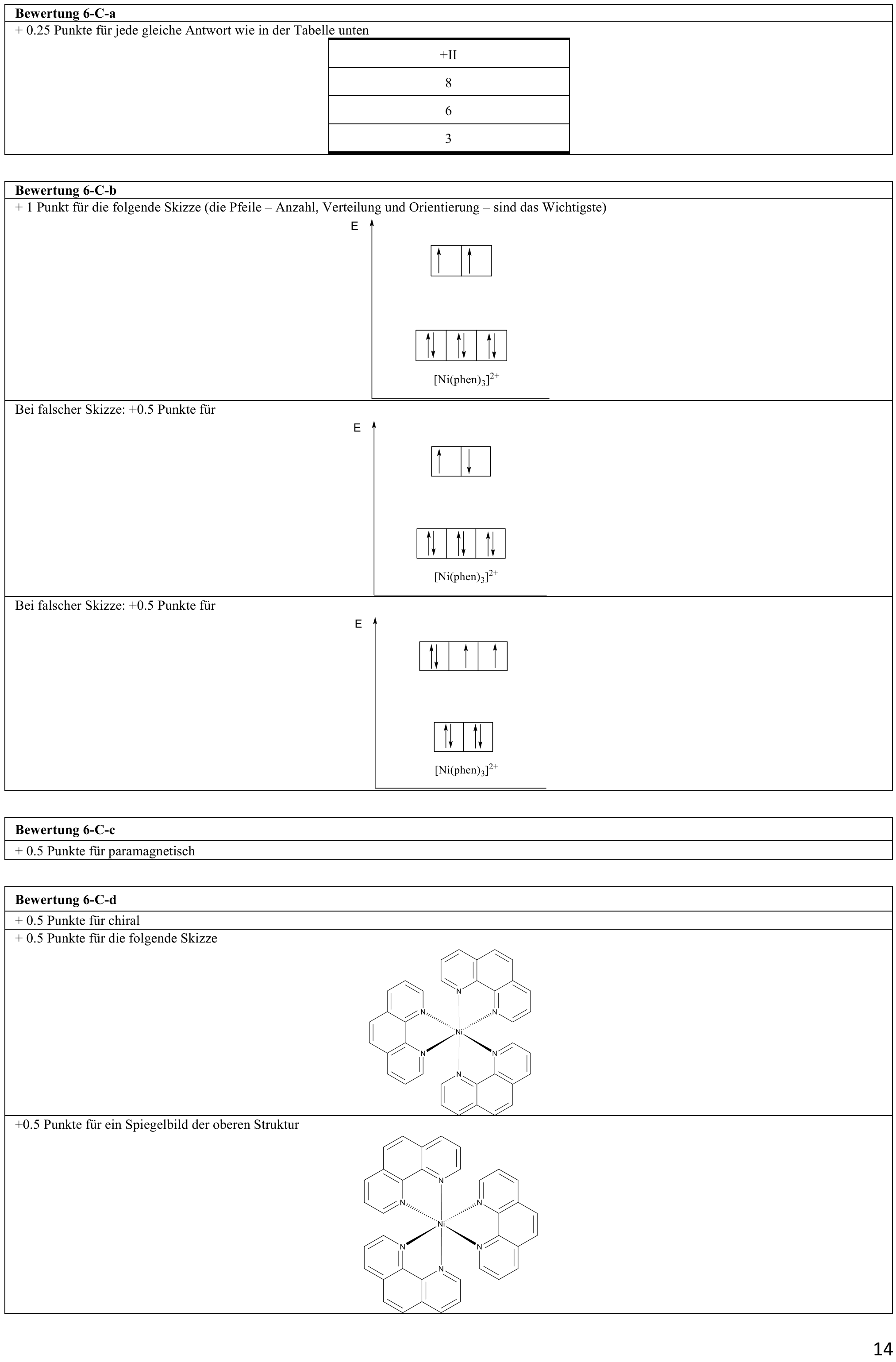}
  \caption{\ The rubric page  associated with the student work in Fig.~\ref{fig:examples}. The instructions are: +0.25 points for each identical answer to the table below. +1 point for following sketch (the arrows --- number, distribution, and orientation --- are most important; if the sketch is incorrect: +0.5 points for \ldots (repeated for both configurations). +0.5 points for paramagnetic. +0.5 points for chiral; +0.5 points for the following sketch; +0.5 points for a mirror image of the above sketch.}
  \label{fig:example_rubric}
\end{figure}

We considered the final exam of the course, which
evaluated students' comprehension and application of foundational chemical principles. The multi-part problems had various formats (multiple choice, fill-in-the-blank, drawing, and graphing) and were distributed across 15~pages. The left and middle panels of Fig.~\ref{fig:examples}  show examples of student work, Fig.~\ref{fig:example_rubric} the associated rubric page.

Each of the six problems addressed a distinct area of introductory general chemistry:
\begin{itemize}
\item The first problem primarily tested general concepts related to atoms, molecules and ions including analysis of the composition of a particle ($^{23}\text{Na}^+$), prediction of VSEPR-structure of molecules or ions ($\text{SeF}_4$ and $[\text{ClO}_3 ]^-$) or calculation of the lattice energy of an ionic compound (NaH) based on provided thermodynamic data. Students had to deal with simple stoichiometric tasks and formulate reaction equations. The problem had 8~parts and was worth 12~points.
\item The second problem focused on the one hand on physical properties of liquids and gases, and students were asked to determine the boiling point, vapor pressure and enthalpy of vaporization of methanol based on graphical data. On the other hand, they analyzed systems, which reached thermal or chemical equilibrium. The former included the determination of the final temperature, the latter the calculation of reaction entropy, Gibbs free enthalpy of reaction or equilibrium constant as well as prediction of the effects of changing temperature and volume on chemical equilibria. The problem had 9 parts and was worth 13~points.
\item The third problem addressed kinetics of chemical reactions and radioactive decays. Students needed to identify isotopes, calculate decay constants, half-lives, reaction rates, and activation energies using provided experimental data. The problem had 6~parts and was worth 8~points.
\item The fourth problem centered around acid-base and solubility equilibria. The students were asked to analyze and complete speciation diagrams and used these to determine pH and p$K_a$ values, concentration of species or solubility products. The problem had 10~parts and was worth 10~points.
\item The fifth problem focused on redox reactions and electrochemistry, requiring students to balance complex redox equations, calculate electrode potentials, and evaluate changes in cell voltage under different conditions. The problem had 6~parts and was worth 7~points.
\item Finally, the sixth problem tested the students' understanding of the relationships between structure and properties. It also included a rather complex stoichiometric task related to mixing solutions of sulfuric acid with different concentrations. Furthermore, the students assessed different aspects of coordination compounds ($(\text{Ni(phen)}_3 )^{2+}$ and $(\text{Ni(H$_2$O)}_6 )^{2+}$), including predicting molecular geometries, electron configurations or magnetic properties. They also examined their chirality and color based on ligand-field theory. The problem had 7~parts and was worth 10~points.
\end{itemize}

\begin{figure*}
\centering
\begin{tabular}{p{0.49\textwidth}p{0.49\textwidth}}
Drawing:
\newline
  \includegraphics[width=0.48\textwidth]{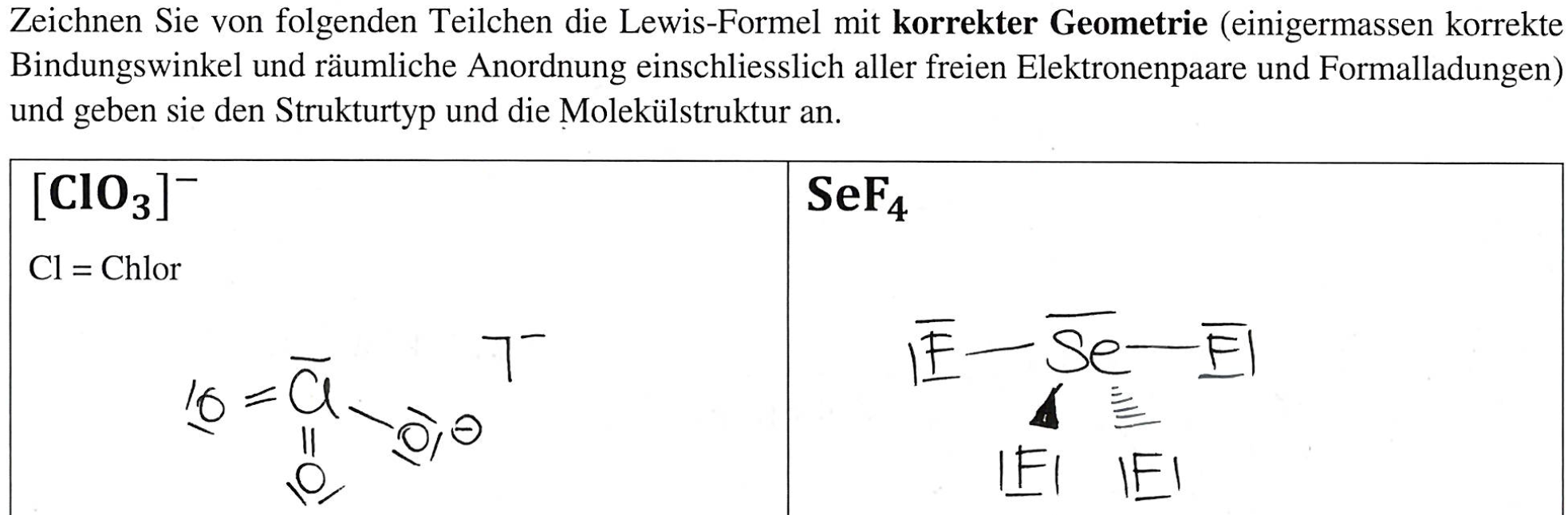}
 \newline 
 
 Graphing:
 \newline
    \includegraphics[width=0.48\textwidth]{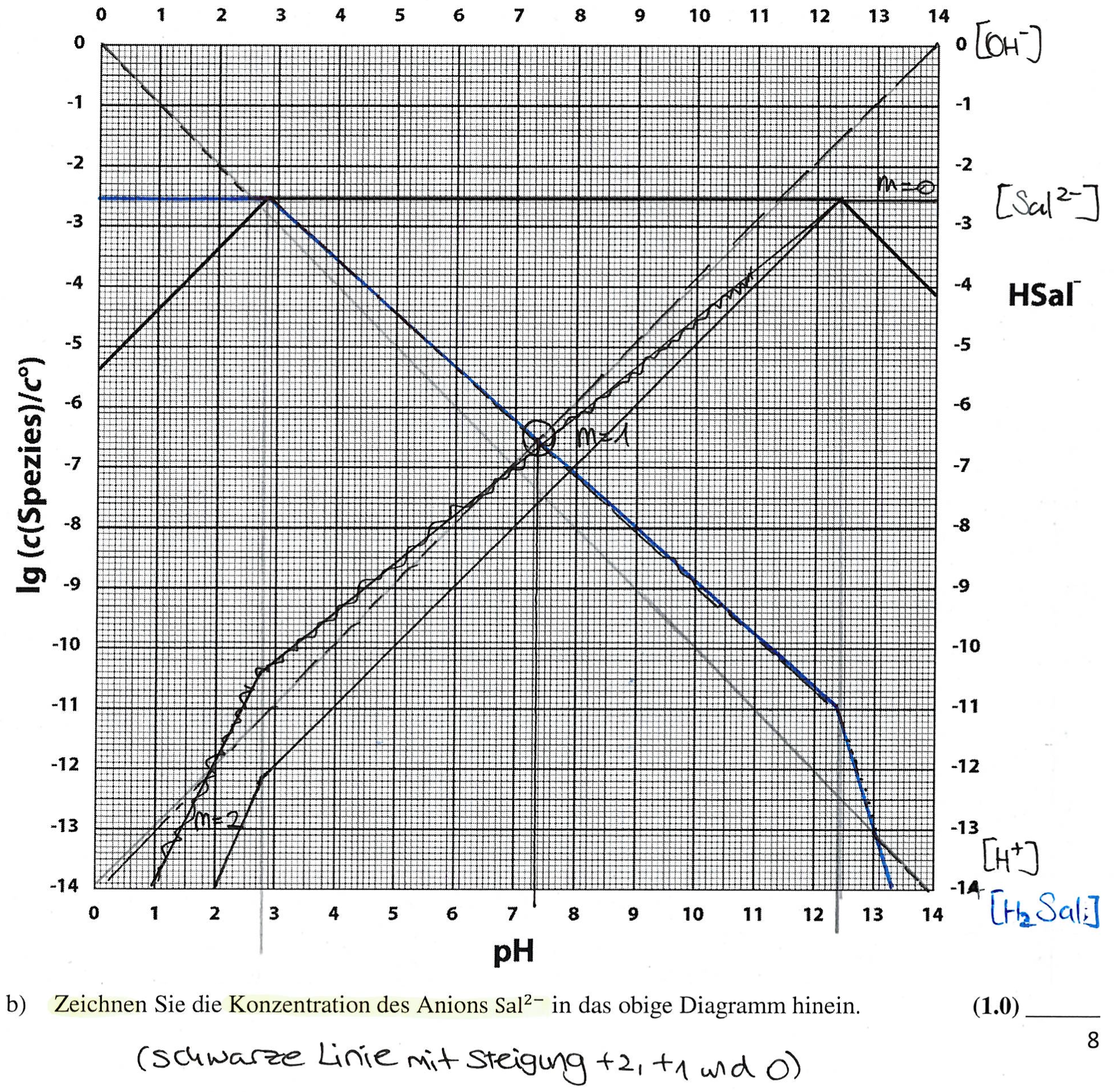}
 \newline
 
    Long answer:
   \newline
    \includegraphics[width=0.48\textwidth]{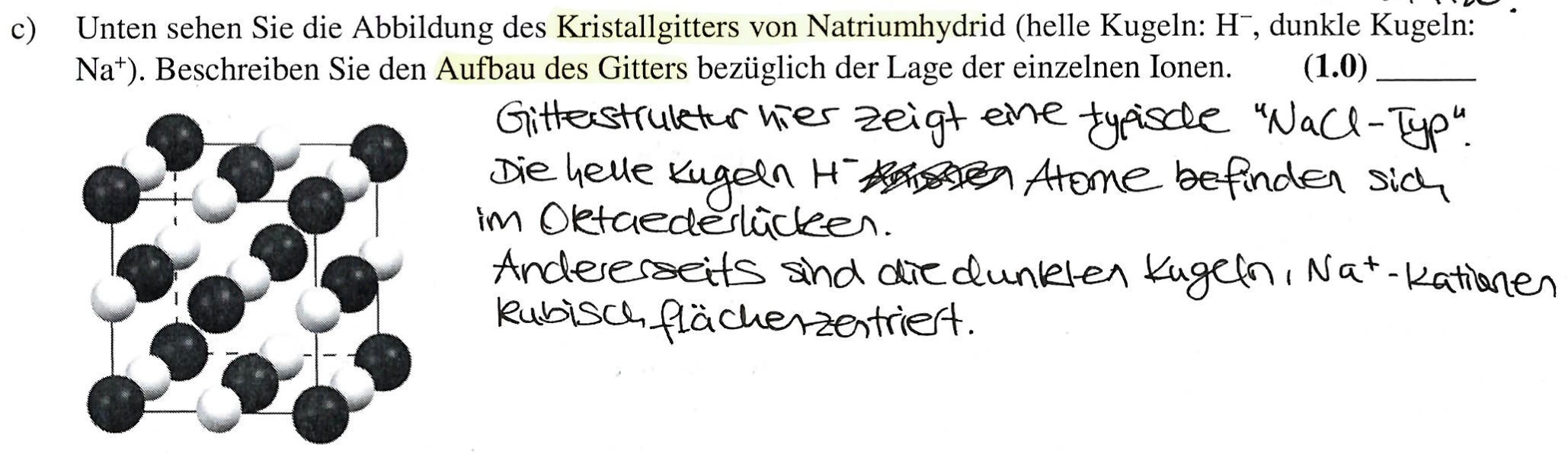}     
&

      Multiple choice:
   \newline
    \includegraphics[width=0.48\textwidth]{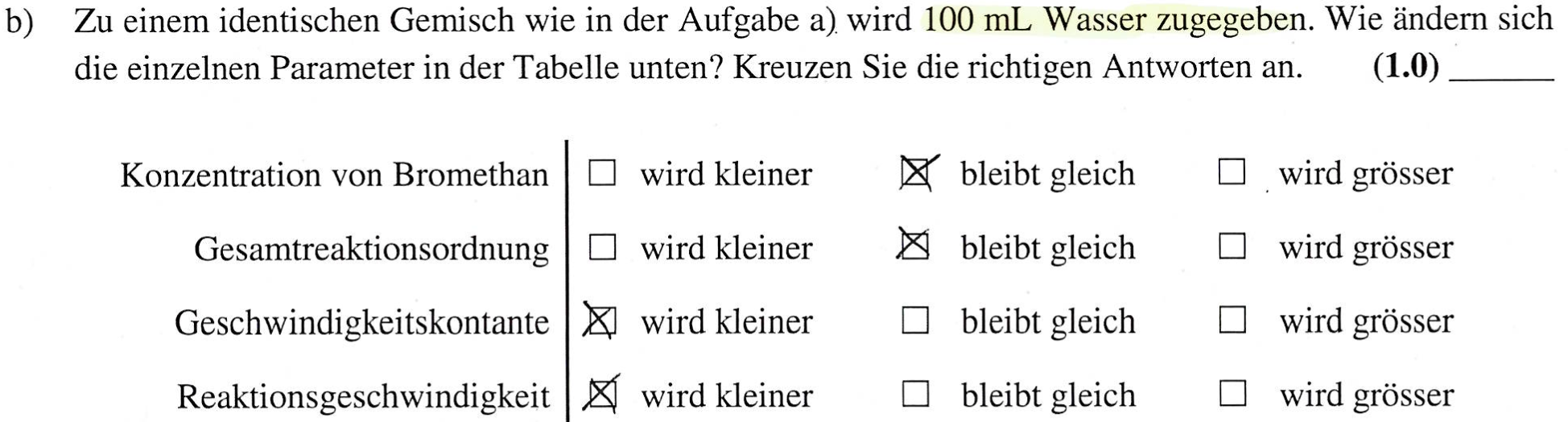}  
\newline

      Numerical:
   \newline
    \includegraphics[width=0.48\textwidth]{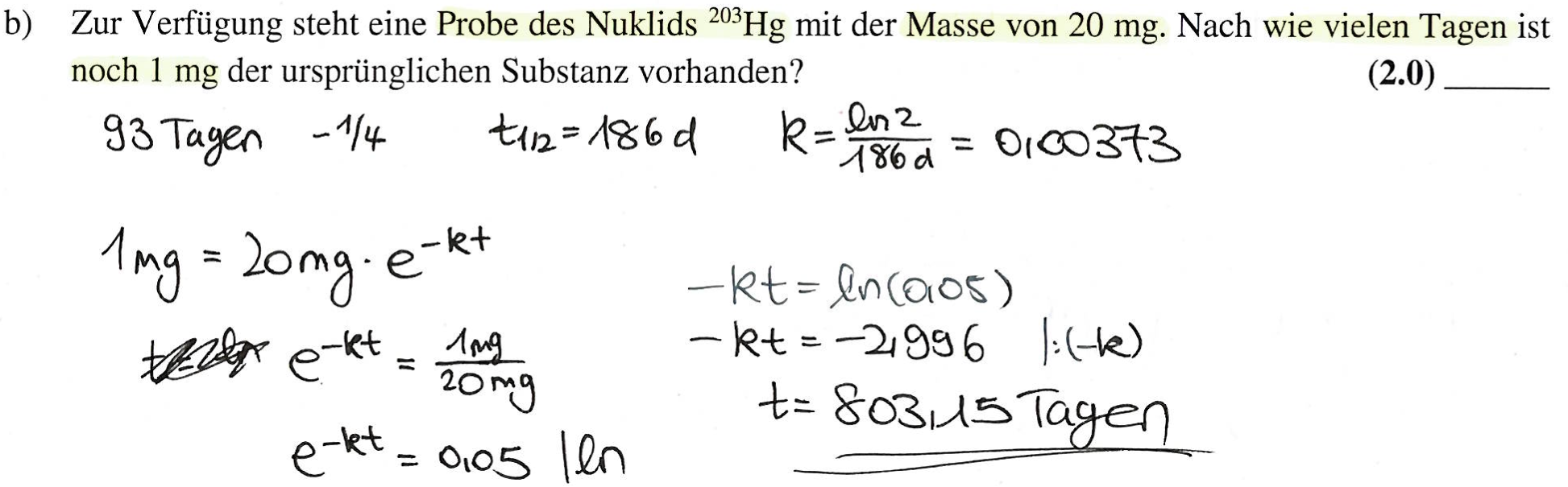}  
       \newline
       
      Reaction:
   \newline
    \includegraphics[width=0.48\textwidth]{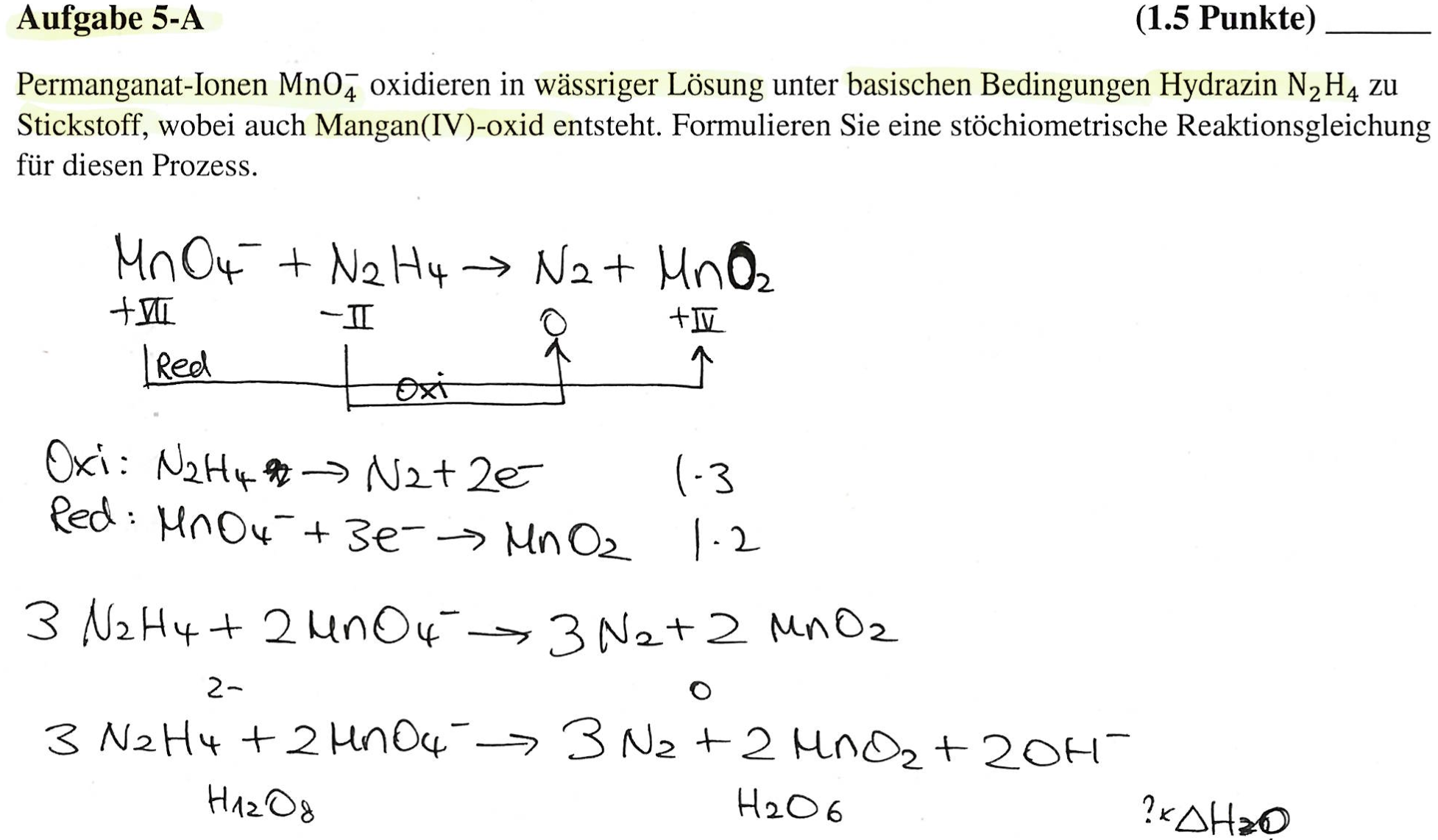} 
              \newline
              
      Short answer:
   \newline
    \includegraphics[width=0.48\textwidth]{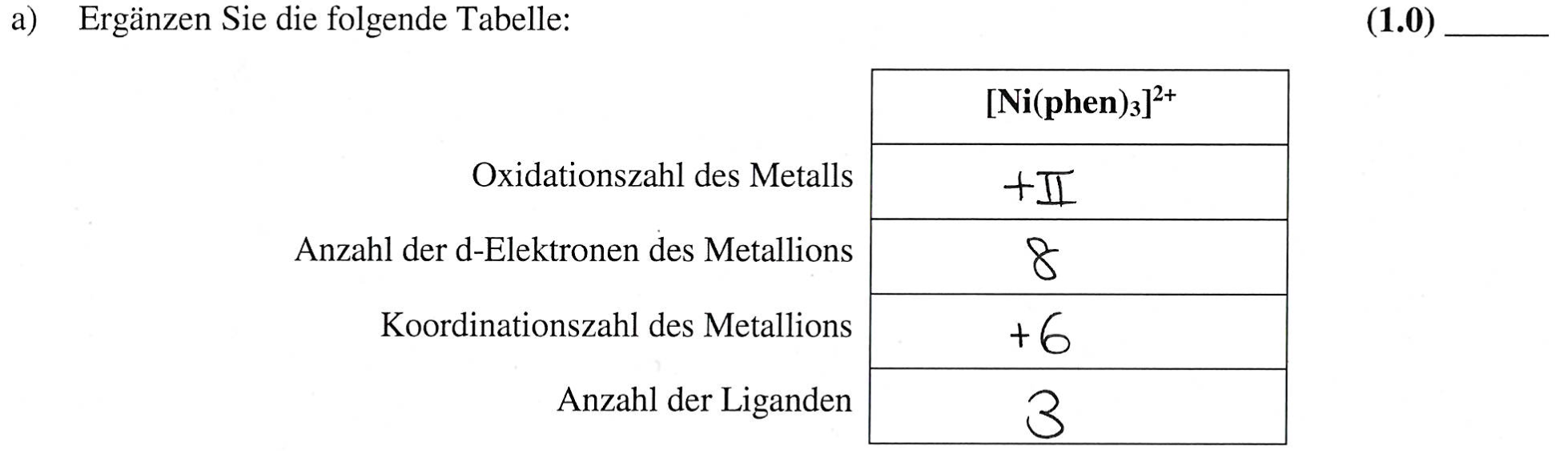} 
          \newline
          
      Symbolic:
   \newline
    \includegraphics[width=0.48\textwidth]{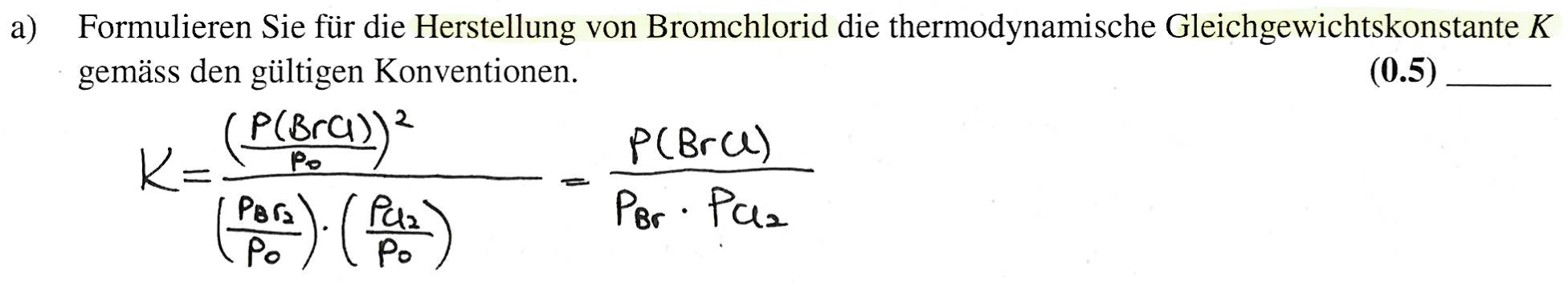}       
    \end{tabular}
  \caption{\ Examples of the problem types on the exam. For the drawing problem: draw the Lewis formula of the following particles with correct geometry (halfway correct binding angles and spatial structure including all free electron pairs and formal charges), and provide the structure type and molecular structure. For the graphing problem: draw the concentration of the anion $\mbox{SaI}^{2-}$ into the above diagram; the student wrote, ``black line with slope +2, +1, and 0.'' For the long answer: below you see the crystal structure of sodium hydride (light spheres: $\mbox{H}^-$, dark spheres $\mbox{Na}^+$). Describe the lattice arrangement with respect to the position of the individual ions;'' the student wrote, ``lattice structure reflects a typical NaCl type, the light spheres $\mbox{H}^-$ are located in the  octahedral voids. On the other hand, the dark spheres, $\mbox{Na}^+$ cations, are cubic face-centered.'' For the multiple-choice problem, the scenario is, ``to an identical mixture like in problem part a), 100~ml of water are added. How do the parameters in the table below change? Mark the correct answer.'' The choices are ``decreases,'' ``stays the same,'' and ``increases'' for each of the concentration of bromoethane, the overall reaction order, the rate constant, and the reaction rate. For the numerical problem: you have a sample of the nuclide $^{203}\mbox{Hg}$ with a mass of 20~mg. After how many days is only 1~mg of the original substance left? In the student response, ``tagen'' means days. For the reaction response: Permanganate ions $\mbox{MnO}_4^-$ oxidize hydrazine $\mbox{N}_2\mbox{H}_4$ to nitrogen in aqueous solution under basic conditions, forming manganese(IV) oxide. Formulate the stoichiometric reaction equation for this process. The short answer problem has already been discussed in Fig.~\ref{fig:examples}. The symbolic problem reads, formulate the thermodynamic equilibrium constant $K$ for the formation of bromine chloride in accordance with the standard conventions.
}  \label{fig:types}
\end{figure*}

Figure~\ref{fig:types} shows the problem types that appeared on the exam:
\begin{description}
\item[Drawing:] Students need to draw a chemical structure.
\item[Graphing:] Students need to draw lines or points into graphs.
\item[Long answer:] Students need to provide an explanation in essay form.
\item[Multiple choice:] Students need to select between different options.
\item[Numerical:] Students need to calculate values.
\item[Reaction:] Students need to write out chemical reactions.
\item[Short answer:] Students need to provide free-form answers.
\item[Symbolic:] Students need to express mathematical relationships in symbolic form.
\end{description}
Some problem parts had multiple types of responses, for example, in one of the problem parts, the students needed to draw a molecule and then name it; we labelled these problem parts ``Combination.''

Overall, the exam required students to demonstrate their ability to integrate conceptual understanding with quantitative problem-solving skills across a broad range of chemical topics. Students could reach up to 60~points across 46~parts.

\subsection{Sample}
A total of 296~students out of 459~students agreed to participate in our study by signing an informed consent. The sample thus included 4,440~pages and 13,616~problem parts. It is important to note that our study was conducted completely in parallel to the traditional grading; all students in the course only received TA-assigned points.

\subsection{Ground Truth}
In this study, we treat the final human-assigned points as the ground truth, as it was produced by the course's routine grading workflow. All solutions were graded by teaching assistants (TAs) under instructor supervision. To maximize within-item consistency, each problem or problem part was graded by a single responsible TA (``one-rater-per-item'' design). Before independent scoring, the instructor and the TA co-graded approximately the first twenty solutions for that item, using a model solution with a detailed point allocation. An exceptions log documented adjudicated alternative solution paths and edge cases so that recurring situations received the same points; during the remainder of scoring, any novel cases were discussed with the instructor and added to the log. The instructor then conducted a brief audit of all graded exams. During the official post-exam review period, students could contest scores; disputed items were re-evaluated by the instructor and any corrections were entered before data export. However, only about one-third of the students took advantage of inspecting the grading, and less than a handful of incorrect grading decisions were discovered. In total, 16 TAs contributed five hours each (about 80 person-hours). Arguably, in spite of all due diligence, mistakes may have been made during grading and slipping through the control mechanisms, and arguably, one might refer to our ground truth as an operational reference standard, but it is that standard that AI-grading would held to.
%
%
\section{Methodology}
\subsection{Data collection}
For students who chose to participate in our study, we scanned their 15-page exam packages into PDF using a standard copier/scanner with automated paper feed. While our contractual framework with the inference provider, Azure AI Services~\cite{azure}, is restrictive enough to permit processing personalized data, our IRB-approved research protocol did not allow this. We thus used a depersonalization scheme with numerical identifiers and a key that was only available to course personnel. The IRB protocol also stipulated that grading personnel were to be kept unaware which students signed the informed consent form to participate in the study, and which students did not.

This depersonalization and scanning process took two individuals a little more than half a day. The pages of the PDFs were subsequently turned into individual PNGs for each page and stored on a server. As it turned out, a handful of students had added a 16th~sheet with additional explanations about certain problem parts; in accordance with the communicated exam policy, additional sheets were neither considered in the TA nor the AI grading of the exam.

After grading was complete, we entered the individual scores for each student and problem part into a spreadsheet to serve as ground truth. Students only had access to their TA-assigned scores; the AI-grading was treated as a parallel experiment.

\subsection{AI systems used for grading}
We used OpenAI models through Azure AI Services~\cite{azure}. The contractual framework stipulates that data is processed on European data centers and that data is not used for training purposes. In particular, we used the multimodal, reasoning model  gpt-o4-mini/high-vision~\cite{gpto4} (model version 2025-04-16).

Each submission to the system included an image of a page of student work (see Fig.~\ref{fig:examples}), as well as an image of the corresponding page from the rubric (see Fig.~\ref{fig:example_rubric}). In other words, we graded page-by-page, always one page at a time (the above-mentioned extra sheets that students turned in could not have been accommodated in this mechanism). The page images were made available to the model from our server via ephemeral URLs (i.e., targeted randomized web addresses, two at a time, that we made accessible only for the duration of one transaction), as embedding them in the submission would have overloaded the $200,000$-token context-buffer of the model.

Fig.~\ref{fig:prompt} shows the prompt used, which produced structured JSON output; this format is easy to process in subsequent steps. Each submission took between 50~and 60~seconds. At times we ran 30~agents in parallel, allowing us to grade the exams in about three hours. We used 32.15 million tokens, costing approximately \$100 in total; tokens are tiny chunks of text (roughly words or pieces of words) that the AI reads and writes, and because Azure charges per token, more text in prompts/context/responses means more tokens --- and thus higher grading cost. This compares to $16\mbox{TAs} \times 5\mbox{h} \times \$44/\mbox{h} \approx \$3,500$ for grading the whole exam for all 459~students (USD~44.00 is the ETH-stipulated gross hourly income of CHF~30.70 plus employer contributions for pensions and insurance of about 15\% at a 2025-exchange rate of $1.25$~USD/CHF).

\begin{figure*}
\centering\small
\begin{verbatim}
        system_prompt = """You are GPT-o4-mini, a high-depth reasoning model with full multimodal (image + text) capability.
A user will send you exactly two PNG images via `image_url` messages:
1. student_image_url - the student's exam page (handwritten + sketches).
2. rubric_image_url  - the corresponding rubric page (tables + partial-credit rules).

Your task:
1. Fetch and visually parse both images.
2. Match each "Bewertung ..." block in the student page to its rubric entry.
3. Apply full?credit and any "Bei falscher  ..." partial-credit rules.
4. Return *only* a JSON array of objects, one per problem_part:
   {
     "problem_part": "6-C-a",
     "awarded_points": 0.75,
     "explanation": "... concise justification ..."
   }

Do **not** output any other text or commentary - *only* the JSON array."""
"""

        messages = [
            {"role": "system", "content": system_prompt},
            {
                "role": "user",
                "content": [
                    {"type": "text", "text": "Please grade this student page using the rubric image below."},
                    {"type": "image_url", "image_url": {"url": stu_url, "detail": "auto"}},
                    {"type": "image_url", "image_url": {"url": rub_url, "detail": "auto"}}
                ]
            }
        ]
\end{verbatim}
  \caption{\ The prompt and JSON messages used for grading. The variables stu\_url and rub\_url would be replaced by the ephemeral URLs generated by our server.}
  \label{fig:prompt}
\end{figure*}

We found that for 3~out of the 13,616~problem parts, the AI assigned a number of points higher than the maximum available points. We left those three grading errors in the data set to reflect the raw data under the assumption that subsequent filtering steps should be robust against such ``misinterpretations'' of the grading rubric.

An important consideration in this workflow was to model a possible production setting as closely as possible. Grading assistance for an exam must not require cumbersome data preparation or prohibitive amounts of manual steps. To broadly apply AI-grading assistance at scale, the net difference in effort must clearly be negative, and the process must be robust and intuitive enough to require minimal technical support.

\subsection{Inter-run reliability of points}
To assess how stable the AI grader is, we ran it five times independently and treated these runs as if they were five human raters. We then asked two related questions: (i) how consistently does the AI assign points to each rubric cell for each student (item$\times$student level), and (ii) how consistently does it produce the \emph{total} score per student (the quantity that ultimately matters for grading).

Our primary metric is the intraclass correlation coefficient (ICC)~\cite{shrout1979intraclass} in its absolute-agreement, two-way random-effects form: ICC(A,1) for a \emph{single} run and ICC(A,$k$) for the \emph{average} of $k$ runs (here $k=5$). Conceptually, the ICC is the proportion of observed variation that reflects real differences between targets (students or cells) rather than run-to-run fluctuation. ``Absolute agreement'' means that any shift or scaling difference between runs counts against agreement; this is appropriate when exact points, not just rank order, matter. We estimate the ICC with the standard variance-components model (two-way random effects for targets and runs).

From the same model, we also report the within-target repeatability standard deviation $S_w$ (the typical run-to-run scatter for the same target) and the 95\% repeatability coefficient $\mathrm{RC}=1.96\sqrt{2}\,S_w$. $\mathrm{RC}$ is directly interpretable on the point scale: for two independent runs, 95\% of absolute differences for the same target are expected to be at or below $\mathrm{RC}$. In addition, because grading sometimes depends on ordering (e.g., relative rank), we summarize the stability of the ranking of total scores across runs using Kendall's $W$ (0 to 1, with 1 indicating identical rankings)~\cite{kendall1939problem}. Finally, we quantify overall level agreement on totals with Lin's concordance correlation coefficient (CCC)~\cite{lawrence1989concordance}, which combines correlation with penalties for any mean or scale bias between runs. For completeness, we also report the median and IQR of the per-student SD of total scores across runs as an intuitive measure of ``wobble.''

All metrics are computed on the exam's raw point scale, which ultimately determines grades (normalizing items by their maximum points is possible and would express results on a 0--1 scale, but we prioritize the raw scale because it aligns with grading decisions).
\begin{description}
  \item[ICC(A,1) vs.\ ICC(A,$k$):] ICC(A,1) reflects reliability of a single AI run; ICC(A,$k$) shows how reliability would improve if one averaged $k$ runs (larger $k$ typically increases ICC).
  \item[$S_w$ and $\mathrm{RC}$ (points):] $S_w$ is the typical run-to-run SD for the same target. $\mathrm{RC}$ is a 95\% bound on absolute differences between two runs; smaller values indicate tighter repeatability. Because it is on the point scale, $\mathrm{RC}$ can also be expressed as a percentage of the exam maximum.
  \item[Kendall's $W$ (rank stability):] Values near 1 mean the ordering of students by total score is essentially unchanged across runs. Because, in many German-speaking university contexts, pass/fail thresholds or grade boundaries are set after inspecting the score distribution (rather than against a fixed absolute scale), stable rankings are particularly important. 
 \item[Lin's CCC (level agreement):] Values near 1 mean runs not only track each other strongly but also show negligible mean/scale bias (agreement along the identity line). Unlike Kendall's $W$, CCC penalizes both mean and scale differences; for absolute point-scale bounds, see the repeatability metrics $S_w$ and $\mathrm{RC}$.
  \item[Per-student SD across runs:] The median SD (with IQR) gives an intuitive sense of how much individual totals vary across runs.
\end{description}

This approach treats both runs and targets as random effects (standard for ICC), assumes run-to-run errors are approximately mean-zero and only weakly correlated across items, and evaluates both fine-grained stability (item$\times$student cells) and practical grading stability (totals). Variability at the cell level may partially cancel when totals are summed, which is why we report both levels.


\subsection{Quality measure}
The quality of the AI-grading is determined through comparison with TA-grading, which was considered as the ground truth.
\subsubsection{Linear regression}
A general quality measure for the AI-grading decisions is to perform linear regressions of the AI-assigned scores versus the established ground-truth scores. This approach assesses how closely the AI's decisions align with human grading. The resulting linear regressions are characterized by three key parameters: slope, offset, and coefficient of determination ($R^2$).

The slope of the regression line indicates how much the AI's scores change, on average, per unit change in the ground truth. A slope of $1.0$ means that the AI's scores increase in perfect proportion to the ground-truth scores, indicating no systematic over- or underestimation across the range of scores. A slope less than $1.0$ suggests that the AI tends to underestimate high scores (or overestimate low scores), while a slope greater than $1.0$ suggests the opposite.

The offset (or intercept) is the value at which the regression line crosses the vertical axis when the ground-truth score is zero. It represents a systematic bias of the AI: for example, a positive offset indicates that the AI tends to assign higher scores than human graders even when the true score is low, while a negative offset indicates the opposite.

Finally, the coefficient of determination, commonly denoted as $R^2$, measures the proportion of variance in the AI's scores that can be explained by the variance in the ground-truth scores. It ranges from $0$~to~$1$, where $1$~indicates perfect agreement (all points lie exactly on the regression line) and $0$~indicates no linear relationship at all. Higher $R^2$ values therefore indicate stronger alignment between AI and human grading.

Together, these three parameters provide a concise, interpretable summary of the quality and reliability of AI grading decisions relative to the human-established standard.
\subsubsection{Classification performance metrics}
To quantify how reliably the AI assigned the exact same fractional score as the instructor, we treated each possible point value (for example, $0$~points, $0.25$~points, $0.5$~points, and so on) as a separate category $C_i$. For each question type (and for the exam overall), we compared every instance where the AI gave a particular score to the corresponding TA-score and noted whether they matched exactly.

With multiple classes $C_i$, the traditional measures for agreement of the model's judgement and the ground truth are defined on a per-class basis as
\begin{description}
\item[True Positive ($\mbox{TP}_i$)] The number of examples that truly belong to class  $C_i$ and were judged a $C_i$, e.g., a problem part that was judged $0.25$ points by both AI and TAs. For example, when considering the class of problem parts with $0.25$ points, the AI judges it $0.25$~points, and in reality, it is indeed $0.25$~points.
\item[False Positive ($\mbox{FP}_i$)] The number of examples that truly belong to some class other than $C_i$ (e.g., $C_k$, $k\ne i$)  but were judged as $C_i$. For example, when considering the class of problem parts with $0.25$ points, an FP would be a part that the AI judges as $0.25$~points, but that in reality it is $0.5$~points.
\item[False Negative ($\mbox{FN}_i$)]  The number of examples that truly belong to $C_i$ but were judged as belonging to some other class $C_k$, $k\ne i$. For example, when considering the class of problem parts with $0.25$ points, an FN would be a part that the AI judges with $0.5$~points, but that in reality it is $0.25$~points.
\item[True Negative ($\mbox{TN}_i$)]  The number of examples that neither belong to $C_i$ nor were judged as belonging to $C_i$
(i.e., everything else that's correctly classified as not $C_i$). For example, when considering the class of problem parts with $0.25$ points, a TN would be a part that the AI judges $0.5$ or $0.75$~points, but that in reality it is $0.5$~points.
\end{description}

From these comparisons we derived four standard multiclass metrics:
\begin{description}
\item[Accuracy:] the proportion of all scores for which the AI's grade exactly agreed with the TA's.
\item[Precision:] averaged equally across all score values, measuring of the grades the AI assigned, how many were correct.
\item[Recall:] also averaged equally over all score values, measuring of the TA's grades, how many the AI captured correctly.
\item[F1 score:] the harmonic mean of precision and recall, giving a single summary of overall exact-match performance.
\end{description}

The quality of a classifier's predictions is often summarized using scores like F1 and Accuracy. While the exact interpretation depends on the situation, some general guidelines apply: F1-scores above $0.8$ usually indicate strong performance, scores between $0.6$ and $0.8$ are considered acceptable, and scores below $0.5$ suggest the predictions may not be very reliable.
It's important to be cautious when interpreting Accuracy in cases where some classes are much more common than others (so-called imbalanced data). For example, a model that always predicts the most common class could still appear to do well in terms of Accuracy, even though it completely ignores the rarer classes.
To address this, we use macro-averaging, which calculates scores (like Precision, Recall, or F1) for each class separately and then averages them. This ensures that each class is treated equally, regardless of how frequent it is. In this way, macro-averaged metrics provide a better picture of overall fairness and performance across all categories:
\begin{eqnarray*}
\text{Accuracy} &=& \frac{1}{n}
\sum_{i=1}^{n}\frac{\text{TP}_i + \text{TN}_i}{\text{Number of Samples}}\\
\text{Precision}& =&\frac{1}{n}
\sum_{i=1}^{n} \frac{\text{TP}_i}{\text{TP} + \text{FP}_i}\\
\text{Recall}&=&\frac{1}{n}
\sum_{i=1}^{n} \frac{\text{TP}_i}{\text{TP}_i + \text{FN}_i}\\
\text{F1}&=&\frac{1}{n}\sum_{i=1}^{n}
  \frac{2\,\text{TP}_i}{2\,\text{TP}_i + \text{FP}_i + \text{FN}_i}
\end{eqnarray*}

Unfortunately, the values of these quality metrics tend to decrease the more classes $n$ there are. Because each distinct point-value becomes its own ``class'' $C_i$,  increasing the number of classes makes exact?match grading exponentially harder --- there are simply more labels to ``get right.'' In a binary setting (two classes), random or naive agreement already sits at 50\% accuracy; with ten classes, chance accuracy is only 10\%. Likewise, macro-averaged precision and recall give equal weight to every class, so rare or poorly-predicted score-values pull the average down just as much as common ones. If there are more possible grades (say from five up to twenty), even competent predictors would see their overall accuracy, precision, recall and F1 slide toward the $1/n$  baseline (where $n$ is the class count). To compensate, we introduce the normed F1:
\begin{equation}\label{eq:normF1}
\text{normed F1}=\frac{\text{F1}-1/n}{1-1/n}
\end{equation}
This value would be $1$ for perfect agreement, $0$ for random guessing, and negative for ``worse than guessing.''


\subsection{Confidence filters}
Generative AI will always generate \textit{something}. Besides the AI-grading itself, a second, equally important part is establishing confidence in the grading outcome~\cite{kortemeyer2025assessing}. Confidence is different from quality measures, since in a production setting, the ground-truth would be unknown, and we need to establish confidence based on the AI-scores alone. For this, we turn to classical and Bayesian statistics.
\begin{figure}
 \centering
  \includegraphics[width=\columnwidth]{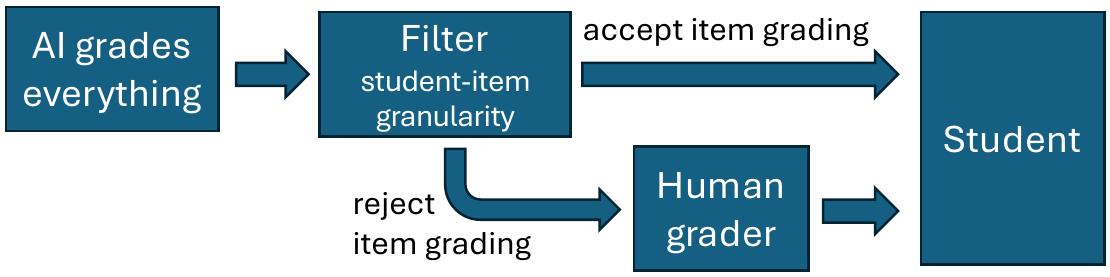}

 \caption{\ Filtering AI-generated student-item grading results.}
 \label{fig:filter}
\end{figure}

Figure~\ref{fig:filter} illustrates this. In a real scenario, there would be no ground truth; instead, initially, only the AI grades all student-items with no human input, generating student-item grades such as ``John Doe, problem part 3a, 2.5 points.'' The confidence filter is used to establish trust or distrust in these AI-judgements. The filter has parameters, which would need to be adjusted solely based on risk tolerance. The stricter the parameters, the more item-gradings will be rejected --- and the higher will be the workload for the humans.
\subsubsection{Partial-credit threshold}
Experiences from earlier studies suggest that AI is more likely to grade correct answers as incorrect (i.e., produce false negatives) that the other way around (false positives)~\cite{kortemeyer2025assessing}. One possible filter is thus to binarize the partial grading decisions and treat negatives as suspicious.
For the purpose of filtering the AI grading decisions, we turn the partial-credit judgements into binary decisions based on a threshold $t$ for ``correctness.'' If a result is ``incorrect" or ``incorrect enough'' based on our threshold, we discard the AI-judgement and, in a production setting, would delegate the grading of that problem part for that student to a human grader.

To assess the quality of these decisions, in addition to linear regressions,
we consider each student $i$'s response to part $j$ as either correct (``positive'') or incorrect (``negative'') based on whether its AI-score $s_{ij}$ meets or exceeds a chosen cutoff $t$.  We then compare the AI's binarized decision to the binarized ground truth and count:

\begin{description}
  \item[True Positive (TP):]  Cases where the AI marks a part as ``correct'' ($s_{ij}\ge t$) and the TA's score also meets or exceeds $t$.  
  \item[False Positive (FP):]  Cases where the AI marks a part as ``correct''' but the TA's score falls below $t$.  
  \item[False Negative (FN):]  Cases where the AI marks a part as ``incorrect'' ($s_{ij}<t$) yet the TA's score meets or exceeds $t$.  
  \item[True Negative (TN):]  Cases where both AI and TAs agree the part is ``incorrect'' ($<t$).  
\end{description}

From these counts we derive the same standard performance measures as for the multiclass metrics, but for only two classes:

\begin{description}
  \item[Accuracy:] the fraction of all parts for which AI and instructor agree.  
    \[
      \text{Accuracy} = \frac{\text{TP} + \text{TN}}
                             {\text{Number of Samples}}.
    \]

  \item[Precision:] of the parts the AI calls ``correct,'' the fraction that truly are correct.
    \[
      \text{Precision} = \frac{\text{TP}}{\text{TP} + \text{FP}}.
    \]

  \item[Recall:] of the parts the instructor called ``correct,'' the fraction the AI also caught.
    \[
      \text{Recall} = \frac{\text{TP}}{\text{TP} + \text{FN}}.
    \]

  \item[F1 Score:] the harmonic mean of precision and recall, balancing those two concerns.
    \[
      \text{F1} = 2 \;\frac{\text{Precision}\ \text{Recall}}
                   {\text{Precision} + \text{Recall}}.
    \]
\end{description}

\subsubsection{Risk threshold}
Psychometrics originated in the early 20th century as a means to standardize the measurement of psychological traits such as intelligence and aptitude. The field introduced techniques like factor analysis to identify latent variables underlying observed responses, reducing them to common dimensions. A prominent early example was the general intelligence factor, $g$, proposed to capture overall cognitive ability~\cite{hart1912general,thurstone1934vectors,spearman1961general}. Over time, such broad constructs gave way to domain-specific abilities shaped by learning, and the focus of psychometrics gradually shifted from evaluating individuals to evaluating the assessments themselves, providing tools to ensure the validity, reliability, and fairness of tests.

Modern psychometric models, including Item Response Theory (IRT), adopt a Bayesian perspective by treating student ability and item difficulty as latent variables inferred from observed responses. These models can be turned around to make predictive statements: given a fitted model, one can estimate the probability that a particular student will correctly solve a particular problem part. In this work, we leverage a two-parameter logistic IRT model to estimate, for each student-part pair, the expected probability of success --- or, in the case of partial credit scoring, a normalized expected value that can serve as a reference against AI-assigned grades.
\[
p_{ij} = \frac{1}{1 + \exp\bigl(-\,a_{j}\,(\theta_{i} - b_{j})\bigr)}\ ,
\]
where $\theta_i$ is the latent ``ability'' of student $i$, and $a_j$ and $b_j$ are the estimated discrimination and difficulty of part $j$, respectively~\cite{rasch,kortemeyer2019quick}.

The AI's actual normalized score $s_{ij}$ is then compared to this expectation~\cite{sinharay2006posterior}, and we define the ``risk'' of accepting that AI judgment as the absolute deviation
\begin{equation}\label{eq:risk}
  \text{Risk}_{ij} = \bigl|\,s_{ij} - p_{ij}\bigr|.
\end{equation}
A risk threshold \(r\) is chosen between 0 and 1, and we accept only those student-part scores for which the AI's estimate lies within $r$ of the model's prediction. In summary:

\begin{description}
  \item[Expected Score ($p_{ij}$):]  The IRT model's probability of student $i$'s earning credit on part $j$, given ability $\theta_i$ and item parameters $(a_j,b_j)$.  
  \item[Observed Score ($s_{ij}$):] The AI-computed normalized partial credit for that student-part.  
  \item[Risk ($\text{Risk}_{ij}$):] The absolute difference between observed and expected, measuring how surprising the AI's judgment is under the Bayesian IRT prior; see Figure~\ref{fig:risk}.  
  \item[Acceptance Criterion:]  Accept if $\text{Risk}_{ij}\le r$; otherwise flag for manual review.  
\end{description}

\subsubsection{Problem types}
Previous experiences have shown that multimodal AI does often succeed in reading handwriting, but still fails to interpret scientific diagrams and graphs~\cite{polverini24,kortemeyer2024grading,kortemeyer2025assessing,kortemeyer2025multilingual}. A straightforward filtering mechanism is thus to exclude any problem parts that have graphical components, namely ``Drawing'' and ``Graphing'' (see Fig.~\ref{fig:types}), i.e., filter for the remaining problem parts that are purely textual or require nothing more than recognizing the position of a cross in multiple-choice problems..

\subsection{Use of AI}
While obviously being the subject of this study, various OpenAI models have also been used for the following aspects of the study: initial drafts of analysis and verification programs in Python, as well as improving the grammar and readability of manuscript passages.

%
%

\section{Results}
\subsection{Inter-run reliability of points}
On \emph{total scores per student}, the AI showed high run-to-run consistency: ICC(A,1)$=0.967$ for a single run and ICC(A,5)$=0.993$ if one were to average five runs. The within-target repeatability SD was $S_w=1.92$ points, implying a 95\% repeatability coefficient $\mathrm{RC}=5.33$ points; i.e., for two independent runs, 95\% of absolute differences in a given student's total would be at or below $5.33$ points on the raw scale. Note that the 95\% repeatability coefficient is a conservative bound; under the same model the \emph{typical} absolute difference between two independent runs is about $2.17$ points ($\approx 3.6\%$ of $60$), and averaging five runs would reduce the 95\% bound to $2.39$ points ($\approx 4.0\%$).
Rank concordance of totals across runs was likewise high (Kendall's $W=0.959$), and level agreement was strong (pairwise Lin's CCC on totals: median $0.977$, IQR $0.045$), indicating both similar ordering and minimal mean/scale bias between runs. As an intuitive dispersion summary, the per-student SD of totals across runs had a median of $1.54$ points (IQR $1.13$).

At the \emph{item$\times$student} level (rubric cells), agreement was lower (as expected for finer granularity), yet still strong: ICC(A,1)$=0.836$ and ICC(A,5)$=0.962$, with $S_w=0.25$ points and $\mathrm{RC}=0.68$ points per cell. The small cell-level jitter helps explain why total scores are very stable: summing across items attenuates unsystematic fluctuations. Overall, these results indicate that single-run totals are highly reproducible, while averaging multiple runs would further reduce residual variation (albeit at added computational and financial cost).

\subsection{Qualitative results}
A comparison of the exam sheets graded by the TAs and score sheets provided by AI, revealed some interesting features of our AI-grading mechanism. This are anecdotal in nature and possibly specific to our grading mechanism (see Sect.~\ref{sec:limitations}).
\subsubsection{False positives}
In some cases, AI assigned points even if the solution was completely wrong. The students were asked to formulate a balanced chemical equation for the formation of sodium hydride from the corresponding elements, according to Eq.~\ref{eq:na}:
\begin{equation}\label{eq:na}
2\ \text{Na}+\text{H}_2 \longrightarrow 2\ \text{NaH }
\end{equation}
AI, however assigned full points also for the equation $2\ \text{Na}^{+}+\text{H}_2^- \longrightarrow 2\ \text{NaH}$,
which is incorrect as the charge balance is disrupted.
Students usually do not complain about receiving points from wrong answers, which means that such false positives would remain undetected post-AI-grading.
\subsubsection{False negatives}\label{sec:fn}
\begin{figure}
 \centering
  \includegraphics[width=\columnwidth]{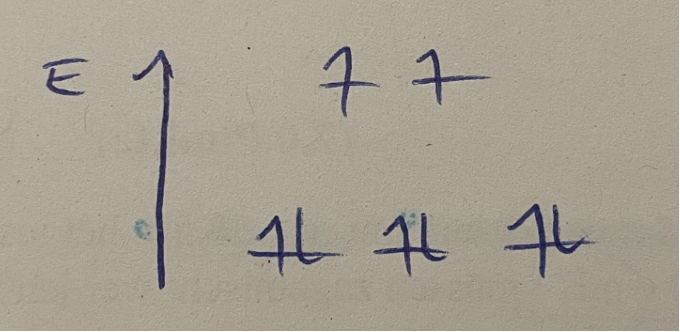}

 \caption{\ An example of a false negative (problem 6-C-b).}
 \label{fig:6cb}
\end{figure}

In terms of false negatives, here are some examples, where AI made an obvious grading error;   this  very often occurred in graphical tasks like the one shown in Fig.~\ref{fig:6cb}.
 
AI misread the sketch. It detected three upward arrows in the higher-energy level and interpreted them as three electrons with identical spin, concluding that the student had indicated an incorrect electron count and therefore assigned zero credit. In fact, the leftmost arrow is not an electron; it is the energy axis of the diagram, labeled $E$. Interpreted correctly, the student's answer is consistent with the rubric, and the TA awarded full credit.

\begin{figure}
 \centering
  \includegraphics[width=\columnwidth]{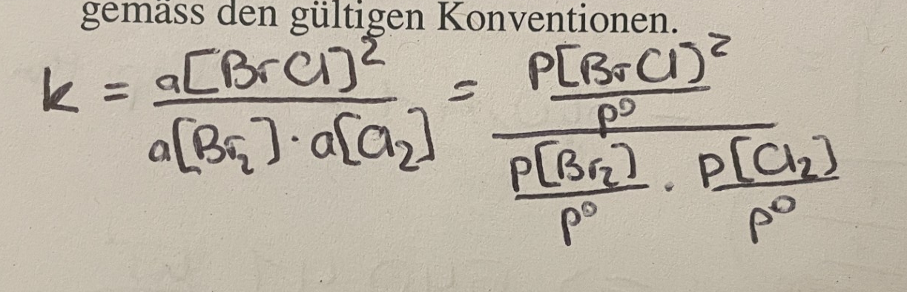}

 \caption{\ An example of a false negative, where AI assigned only partial credit (problem 2-C-a).}
 \label{fig:2ca}
\end{figure}
AI assigns only partial points for the expression with the activities shown in Fig.~\ref{fig:2ca}, as it ignores the subsequent fraction, which contains the expected   partial pressures of the individual gaseous species in the reaction equation. Earlier, we found anecdotal evidence that the incorrect alignment of the fractional line may play a role.
 
\begin{figure}
 \centering
  \includegraphics[width=0.4\columnwidth]{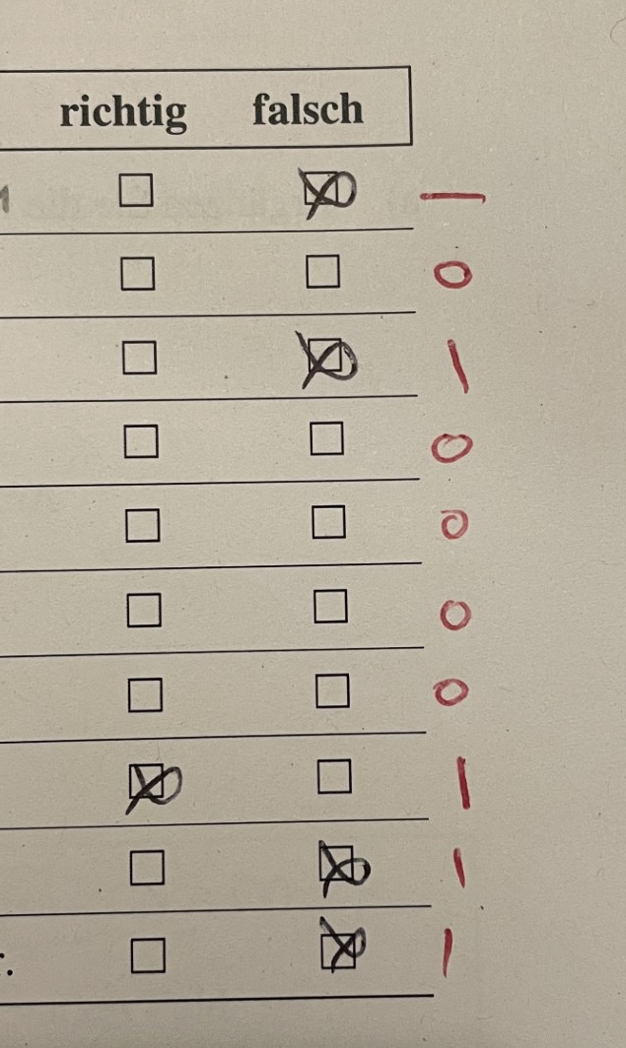}

 \caption{\ Incorrect interpretation of a multiple-response problem (problem 6-A).}
 \label{fig:6a}
\end{figure}
 
In Fig.~\ref{fig:6a}, AI does not correctly recognize the position of a cross. It claims that the second-to-last answer contains the cross in the left column named ``richtig'' (correct) and deducts points. It is unclear how this may have happened, except that the student markings are not really crosses.

\begin{figure}
 \centering
  \includegraphics[width=\columnwidth]{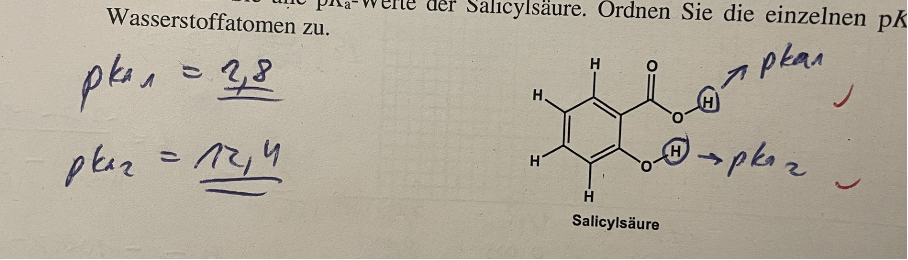}

 \caption{\ Incorrect interpretation of labels (problem 4-A-a).}
 \label{fig:4Aa}
\end{figure}
For the solution shown in Fig.~\ref{fig:4Aa}, 
AI awards only partial points as it claims: ``Correct numeric pKa values (2.8 and 12.4) but assigned to the wrong hydrogen atoms.'' The provided answer of the student is, however, entirely correct.  Very likely, the system did not recognize correctly how the labels $\text{pKa}_1$ and $\text{pKa}_2$ were attached to the graphical representation of the molecule, making this another example of the system struggling with graphics.

In one of the tasks, students were asked to classify several statements as true or false. According to the rubric, correct classifications earned credit and incorrect classifications resulted in a deduction, with the overall score for the item constrained to be $\ge0$. Contrary to the rubric, the AI ignored this zero floor and assigned negative totals for some responses.

\subsubsection{Corner-cases, judgement calls, and false false-negatives}
Some grading differences reflected human discretion. Teaching assistants frequently awarded partial credit when the final answer was incorrect but the work demonstrated a serious attempt --- for example, laying out a sensible approach, articulating reasoning, or pursuing an individual yet defensible method.
Encoding all such contingencies in a rubric for an exam taken by several hundred students would be impractical. Similarly, ``error-propagation'' cases were handled leniently by TAs: when a student carried a miscalculated intermediate from one part to the next but applied the correct strategy thereafter, full credit was typically awarded. TAs were instructed to re-check subsequent calculations using the student's (incorrect) carried-forward value.

At times the AI was stricter (or simply more consistent) than a TA. In a task asking for the product of the $\beta$-decay of ${}^{203}\text{Hg}$, the AI correctly flagged ${}^{203}\text{Ti}$ as incorrect and assigned no points. A teaching assistant likely misread the handwriting --- the small difference between \textit{Ti} and \textit{Tl} --- and interpreted the response as ${}^{203}\text{Tl}$, awarding full credit. While the exact cause cannot be verified, this example highlights a potential advantage of AI-assisted grading: it does not tire and is less vulnerable to perceptual slips.

\subsection{Unfiltered quantitative results}

\begin{figure*}
 \centering
  \includegraphics[width=0.33\textwidth]{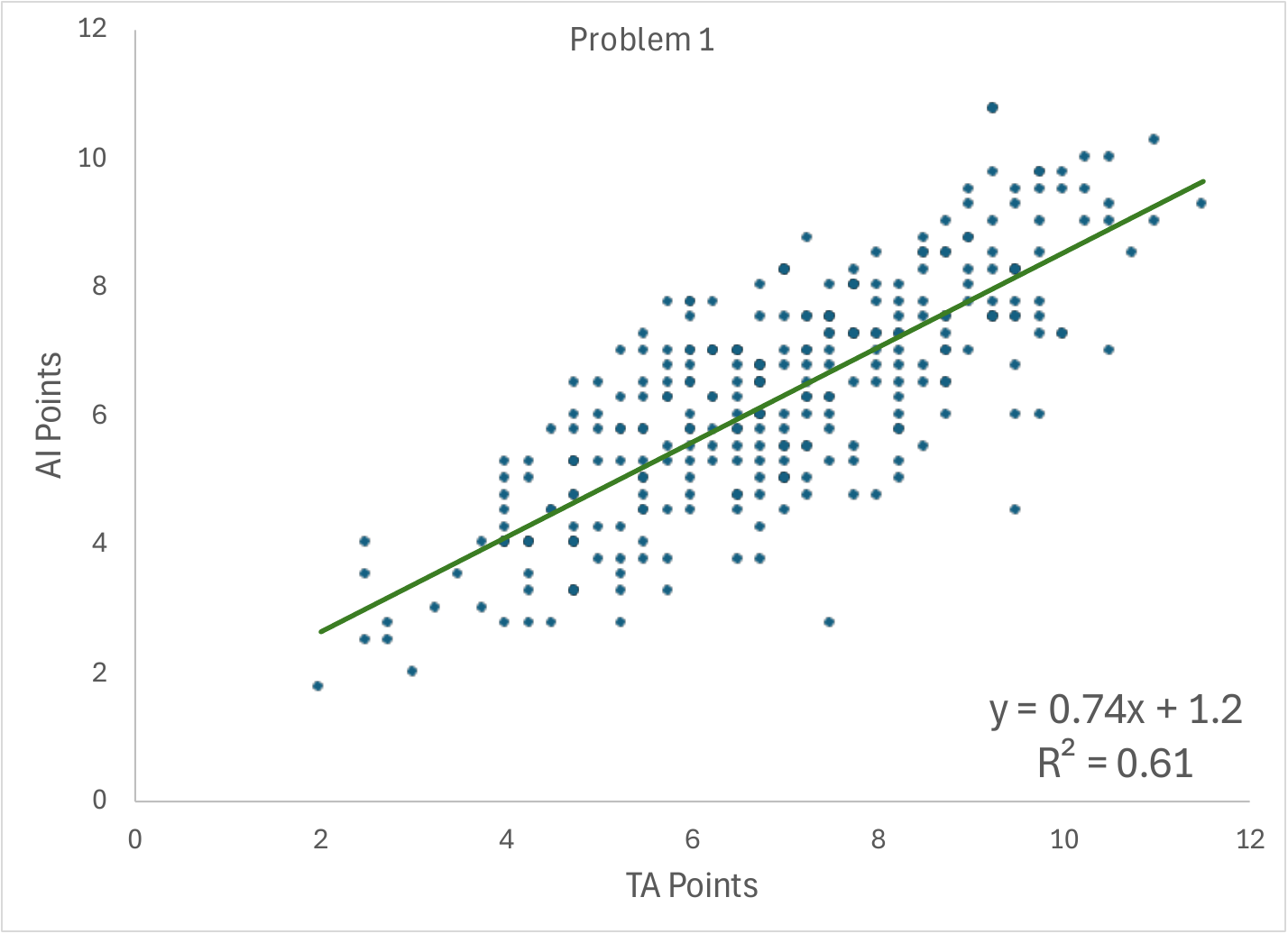}
  \includegraphics[width=0.33\textwidth]{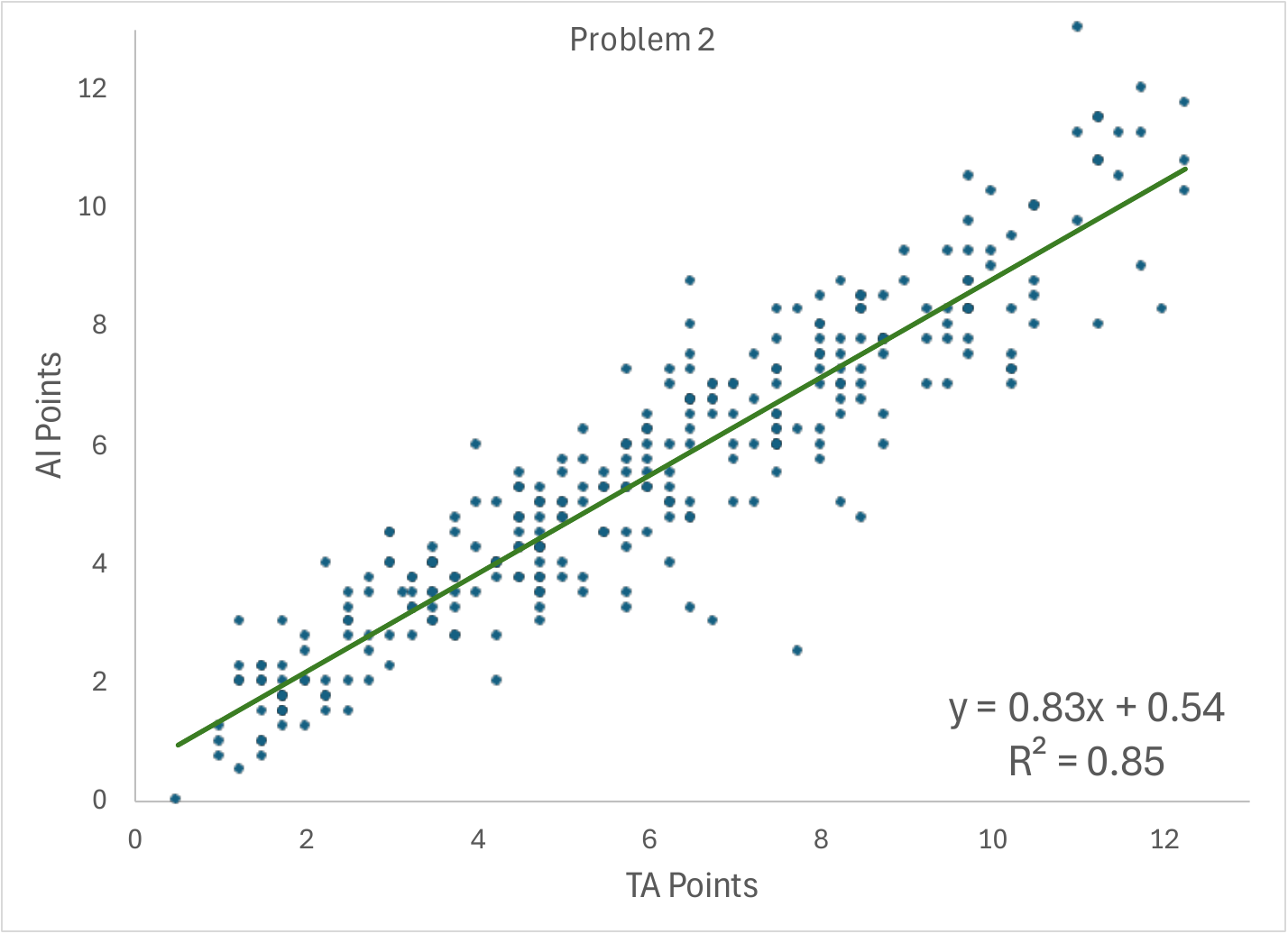}
  \includegraphics[width=0.33\textwidth]{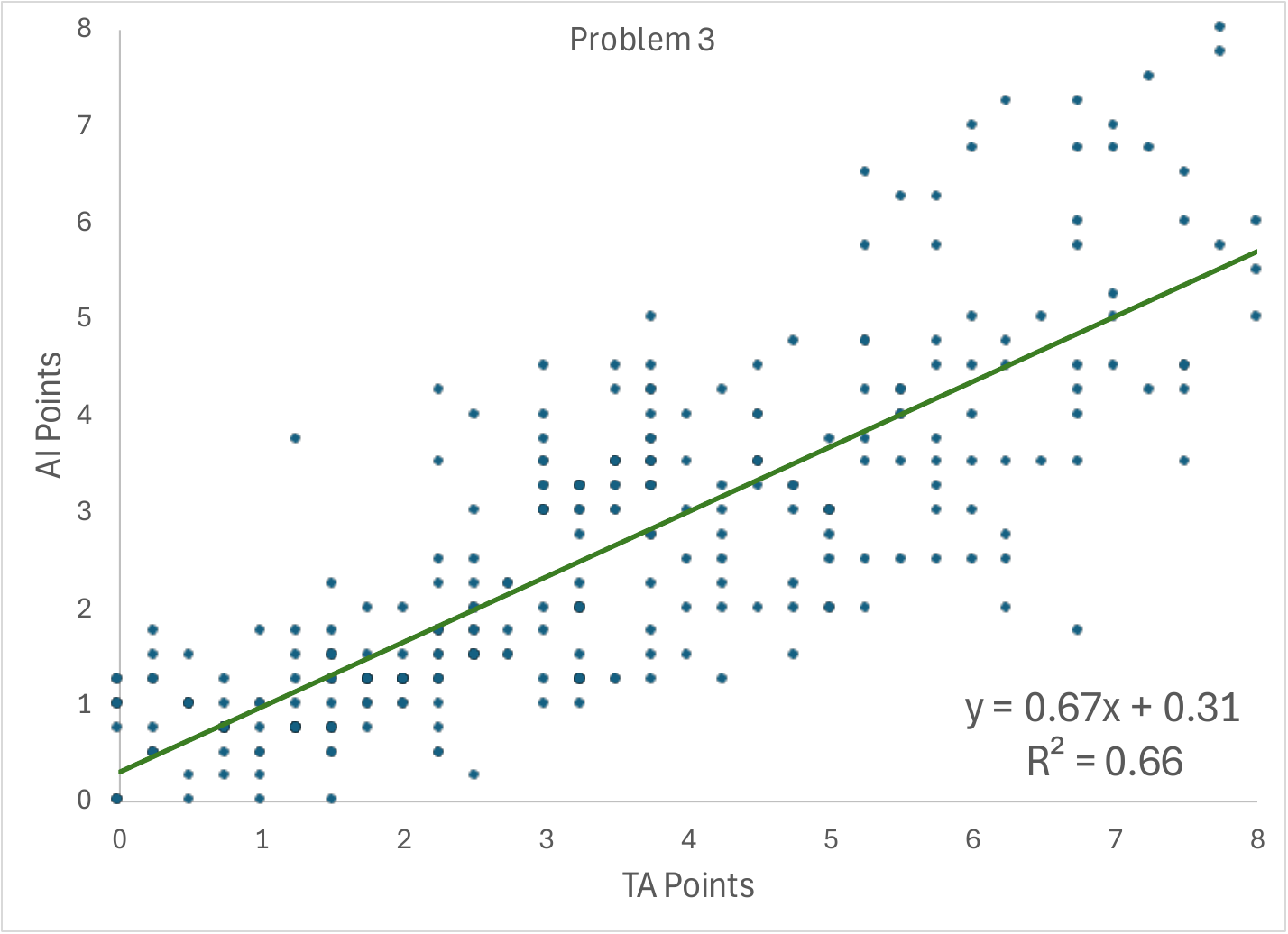}
  
   \includegraphics[width=0.33\textwidth]{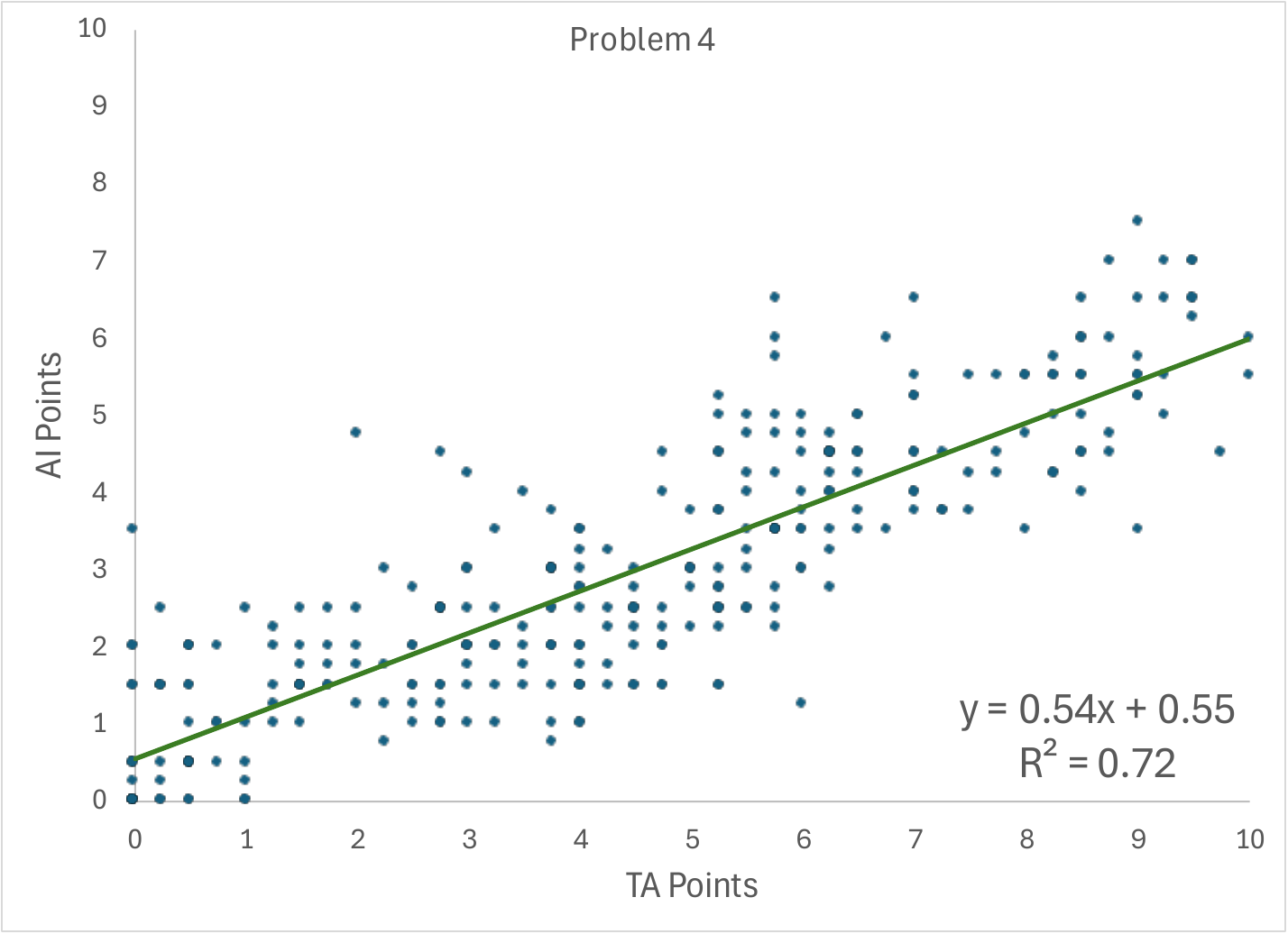}
  \includegraphics[width=0.33\textwidth]{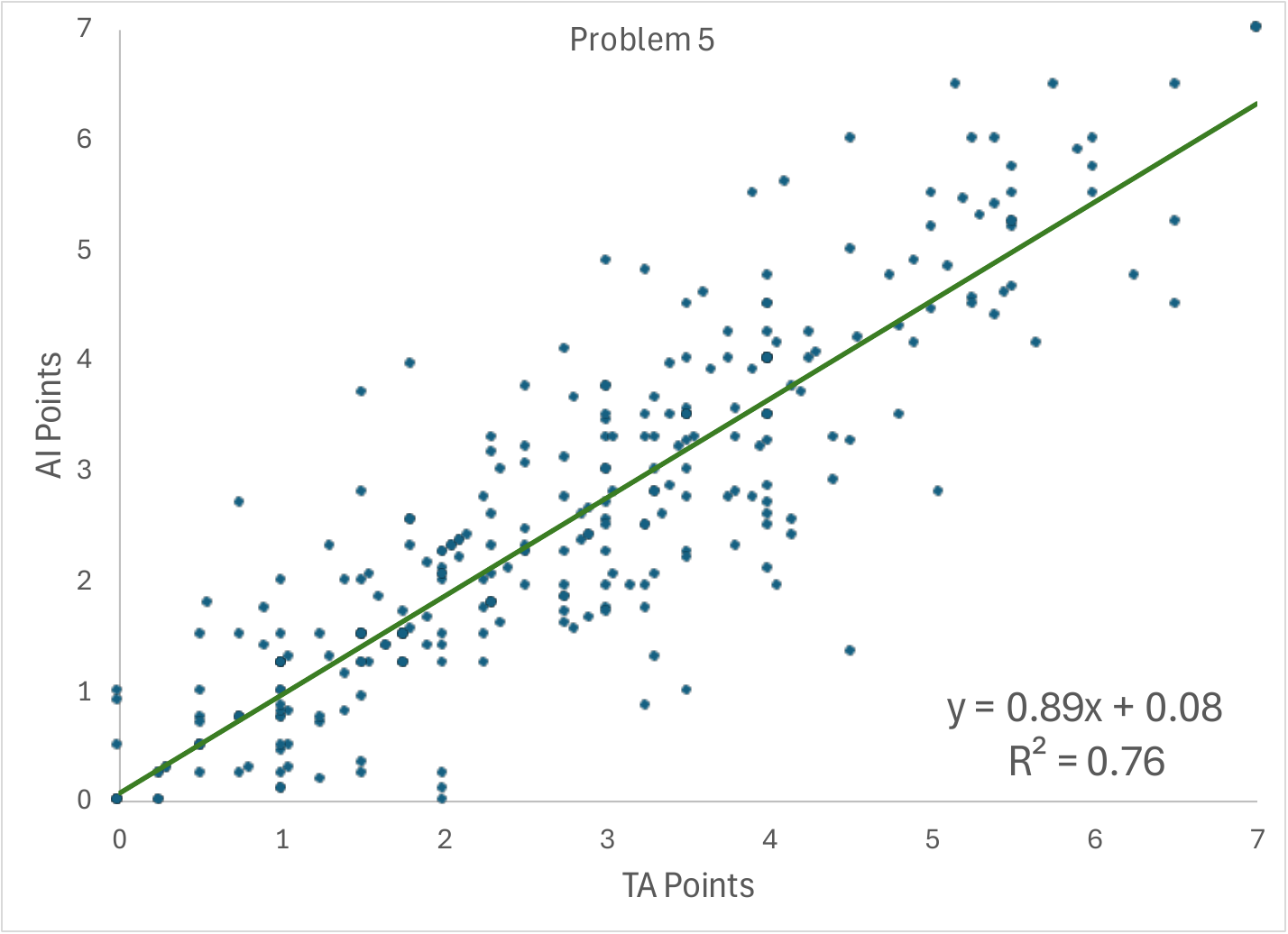}
  \includegraphics[width=0.33\textwidth]{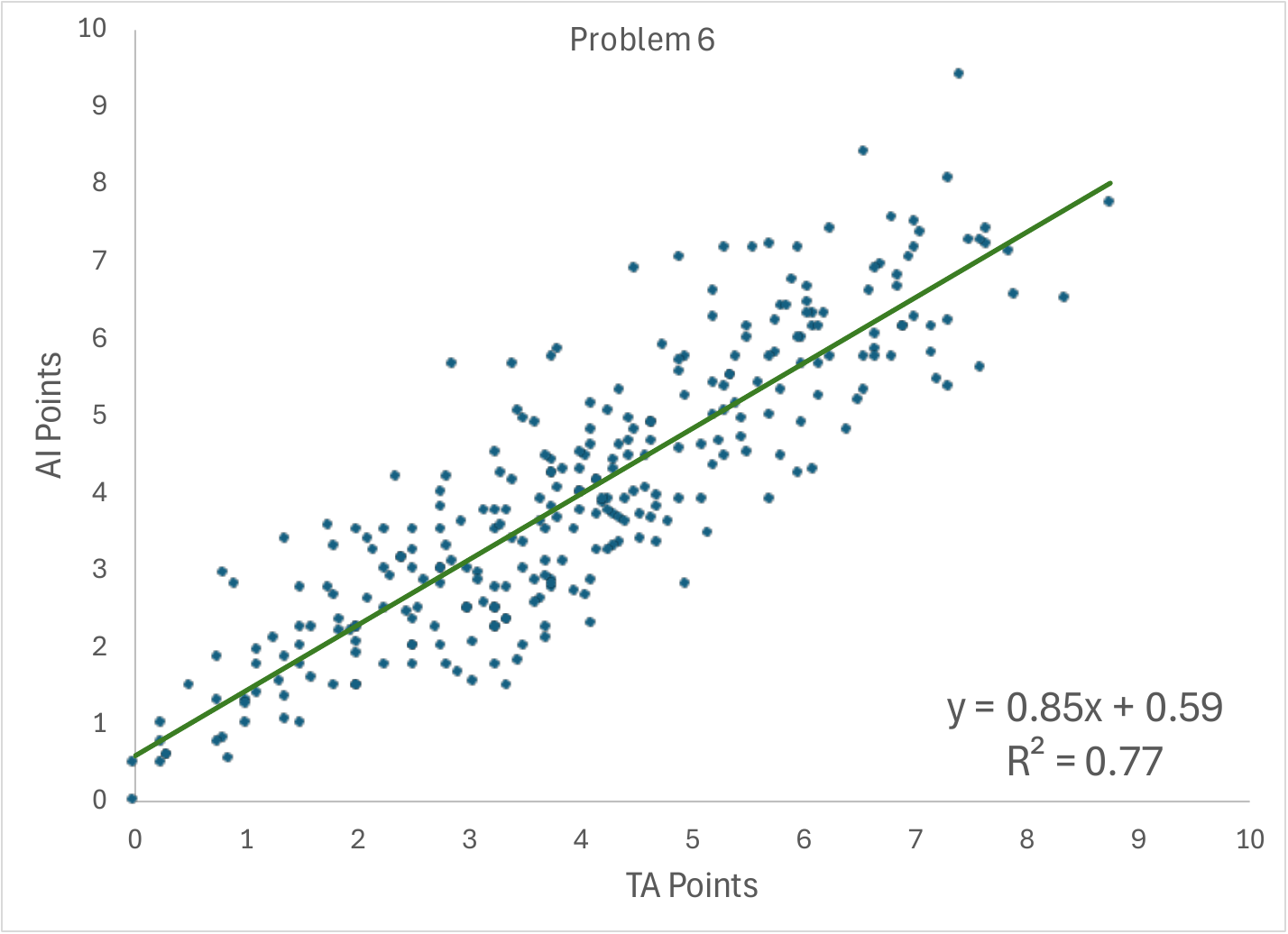}
 \caption{\ Scatterplots and linear regressions of the AI-assigned versus the TA-assigned points for the six problems on the exam. Each data point represents one student.}
 \label{fig:problems}
\end{figure*}

\begin{figure}
 \centering
  \includegraphics[width=0.49\textwidth]{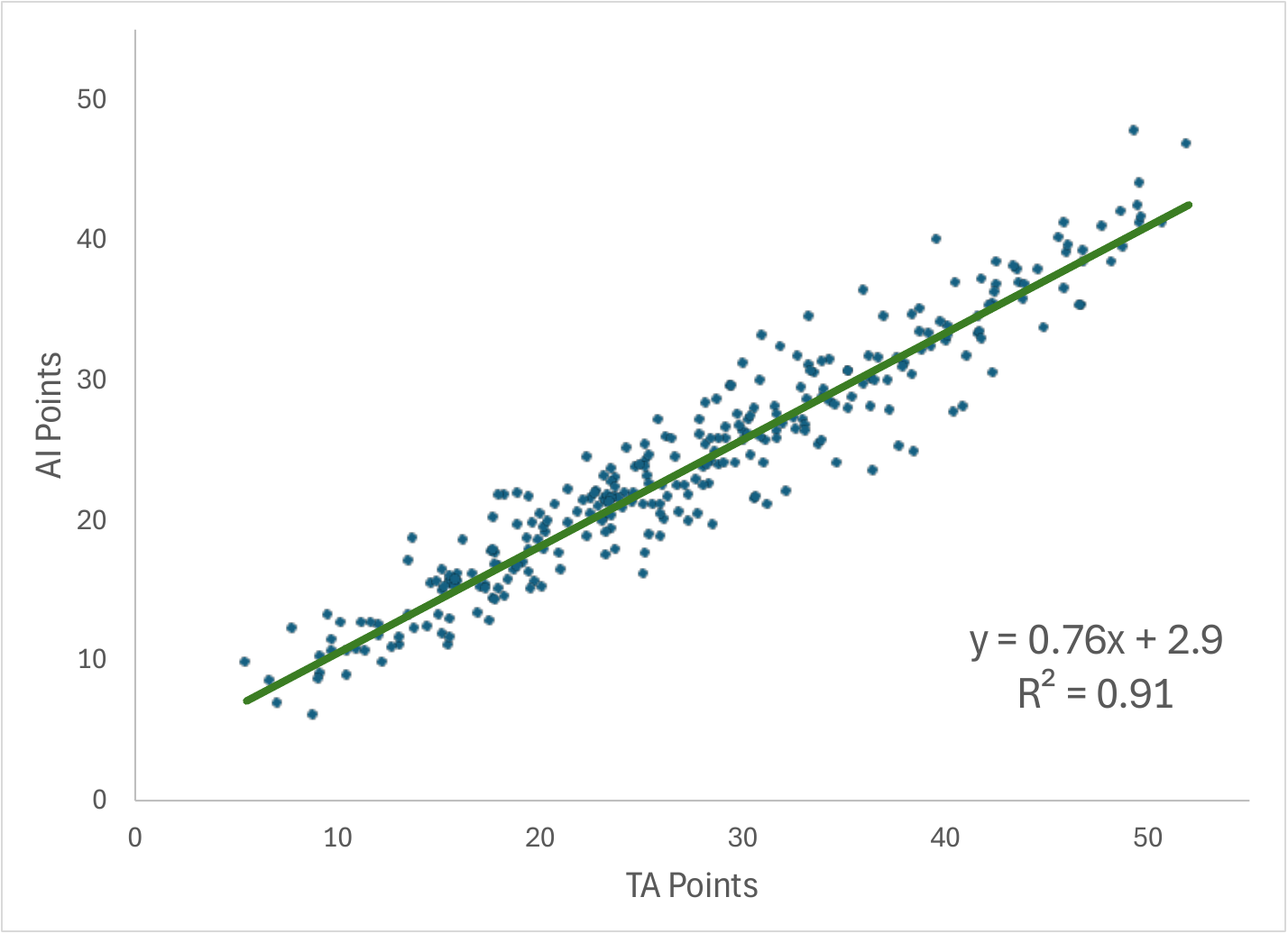}
 \caption{\ Scatterplots and linear regressions of the AI-assigned versus the TA-assigned points for the full exam. The maximum achieved score was~52; as typical for exams in the German-speaking university tradition, no student achieved full points.}
 \label{fig:totals}
\end{figure}

As Fig.~\ref{fig:problems} shows, for the individual six problems, the AI-derived scores correlated with human scores at coefficients of determination ($R^2$) ranging from $0.61$ to $0.85$. The reliability on a per-problem basis is low and not suitable for grading without additional measures, yet the aggregate total score shown in the top-left panel of Fig.~\ref{fig:totals} achieved an $R^2$ of $0.91$ --- that this $R^2$ is ``bigger than the sum of its parts'' may initially appear surprising. The enhancement results from the partial cancellation of random scoring errors across problems: when individual deviations offset one another in the sum, overall noise diminishes relative to the true-score variance. The resulting increase in the signal-to-noise ratio produces a more reliable composite measure of student proficiency than any single item alone. In this context, by ``noise'' we mean unsystematic item-level misgrading --- some items scored too high and others too low --- so aggregation partially cancels these opposite errors, increasing the reliability of the total (as predicted by the Spearman--Brown prophecy in Classical Test Theory). This argument, borne out by our result, assumes that item-level errors are approximately mean-zero and only weakly correlated across problems; systematic biases would not cancel.

As Table~\ref{tab:allqf} shows, the multiclass quality of the grading varies between question types. The first column lists the question type with additional rows across all question types and a row for combined question types.  The second column shows the number of these parts, the third column the number of student-part samples, the fourth column the number of classes $n$, the four following columns the quality measures, and the rightmost column the normed F1 (Equation~\ref{eq:normF1}). The table is sorted in order of decreasing normed F1 to highlight the question types that are most accurately graded.

Overall, these quality measures might appear low, however, one needs to keep in mind that these are multiclass measures. If a problem part is worth two points, and possible points are 0.0, 0.25, \ldots, 2.0, these are nine classes (``notches''), and a true positive would only be counted if the fractional credit agrees {\it exactly}; if the ground truth were 1.25 points and the AI assigned 1.5 points, the AI-grading would be considered just as incorrect as if it had assigned 0.25 points.

\begin{table*}
\small\centering
  \caption{\ Multiclass quality measures for the AI grading results in order of decreasing F1.}
  \label{tab:allqf}
  \begin{tabular*}{0.8\textwidth}{@{\extracolsep{\fill}}lrrrlllll}
    \hline
 Type & Parts & Samples & Classes & Accuracy & Precision & Recall & F1 & normed F1\\
 \hline
Long Answer & 1 & 296 & 5 & $0.8$ & $0.68$ & $0.7$ & $0.68$ & $0.6$ \\
Short Answer & 6 & 1776 & 9 & $0.74$ & $0.64$ & $0.6$ & $0.6$ & $0.55$ \\
Reaction & 3 & 888 & 6 & $0.66$ & $0.62$ & $0.58$ & $0.59$ & $0.5$ \\
Multiple Choice & 9 & 2664 & 20 & $0.69$ & $0.41$ & $0.41$ & $0.4$ & $0.37$ \\
Numerical & 18 & 5328 & 10 & $0.65$ & $0.42$ & $0.41$ & $0.41$ & $0.34$ \\
Symbolic & 3 & 888 & 6 & $0.65$ & $0.49$ & $0.43$ & $0.43$ & $0.31$ \\
\textit{All} & 46 & 13616 & 27 & $0.64$ & $0.31$ & $0.29$ & $0.3$ & $0.27$ \\
\textit{Combination} & 4 & 1184 & 13 & $0.36$ & $0.22$ & $0.22$ & $0.2$ & $0.13$ \\
Drawing & 1 & 296 & 5 & $0.56$ & $0.31$ & $0.37$ & $0.3$ & $0.13$ \\
Graphing & 1 & 296 & 5 & $0.33$ & $0.23$ & $0.21$ & $0.12$ & $-0.098$ \\
\hline
\end{tabular*}
\end{table*}

Not surprisingly for a \textit{language} model, short and long textual answers emerge as the most reliably graded one. Fortunately for chemistry education, equations of chemical reactions are also a member of the most reliably graded class of question types. Not reliably graded are questions that require graphical representations, where students are drawing lines into graphs; here, the system performs worse-than-random. It should be noted that these graphing problems involved a background grid, which is visually distracting for AI-vision as all details of the graphs are encoded regardless of them being foreground or background.

While the raw performance of the model may be acceptable for a low-stakes quiz, it is unacceptable for an exam. As it is unlikely that we can increase grading performance, the next-best strategy is to filter out suspicious grading judgements.


\subsection{Filtering based on partial-credit threshold}\label{sec:pf}
Table~\ref{tab:pf} reports the performance of selective automation for AI-assigned partial credit as a function of the acceptance threshold 
$t$ For each student-part cell, the AI proposes a fractional score; if that score is at least 
$t$, the item is auto-accepted, otherwise it is deferred to human grading. ``Accepted'' gives the resulting coverage (percentage of items handled by the AI). ``MAE'' is the mean absolute error in points computed only over accepted items. ``Under'' and ``Over'' are the percentages of accepted items on which the AI under- or over-credits relative to human grading. ``Under > 0.25 Pts'' flags accepted items that are off by what is generally more than one grading notch (0.25 points) on the under-credit side. ``Bias Pts'' is the mean difference (AI $-$ ground truth, in points) on the accepted set, with negative values indicating a tendency to short students and positive values indicating generosity.

\begin{table*}
\small\centering
  \caption{\ Accepted = share of items auto-graded (AI score $\ge t$); MAE = mean absolute point error on accepted items; Under/Over = share of accepted items where AI is below/above the human score; Under > 0.25 Pts = share of accepted items under-credited by more than 0.25 points; Bias Pts = mean(AI $-$ ground truth) on accepted items.}
  \label{tab:pf}
  \begin{tabular*}{0.8\textwidth}{@{\extracolsep{\fill}}lllllll}
    \hline
Threshold $t$&	Accepted&	MAE Pts&	Under&	Under > 0.25 Pts&	Over	& Bias Pts\\
\hline
0 &	100\% &	0.21	& 24\% &	15\% &	14\% &	-0.08 \\
0.1&	59\% &	0.22	& 21\% &	12\% &	24\% &	0 \\
0.2&	57\% &	0.21	& 20\% &	12\% &	24\% &	0.01 \\
0.3&	52\% &	0.21	& 18\% &	10\% &	24\% &	0.04 \\
0.4&	47\% &	0.18	& 11\% &	6\% &	26\% &	0.08 \\
0.5&	47\% &	0.18	& 11\% &	6\% &	25\% &	0.08 \\
0.6&	36\% &	0.16	& 6\% &	3\% &	26\% &	0.11 \\
0.7&	33\% &	0.14	& 4\% &	1\% &	23\% &	0.12 \\
0.8& 	28\% &	0.13	& 1\% &	0\% &	22\% &	0.12 \\
0.9&	28\% &	0.12	& 0\% &	0\% &	21\% &	0.12 \\
1&	27\% &	0.12	& 0\% &	0\% &	20\% &	0.12  \\
 \hline
  \end{tabular*}
\end{table*}

A threshold $t=0$ means unfiltered.
As the threshold increases, the system becomes more conservative: coverage falls, but the accepted items are cleaner. The MAE is always smaller than a grading notch, and it improves further with increasing threshold. The ``Under'' rate, alongside with the ``Under > 0.25 Pts'' rate contract to zero at high thresholds; in other words, tightening $t$ effectively suppresses the student-harm modes, so both the frequency and the magnitude of under-credit among the auto-graded items decrease. The bias on the accepted set shifts from slightly negative at very low thresholds toward small positive values at higher thresholds, indicating that once we keep only higher-scored cases, the AI is no longer prone to short students on average.

From a grade-integrity and workflow perspective, the residual ``Over'' rate on the accepted set remains a salient consideration: even when student-harm is low, some accepted items may be graded a bit generously. The table therefore supports policy choices by making the trade-off explicit: one can select the highest threshold that keeps ``Under > 0.25 Pts'' below a tolerance deemed acceptable (e.g., 1$-$5\%) while still delivering useful coverage. In practice, this yields a defensible operating point that protects students against under-credit on auto-graded work, limits systematic bias, and preserves human review for the low-score and borderline cases where it adds the most value.

\begin{figure*}
 \centering
  \includegraphics[width=0.33\textwidth]{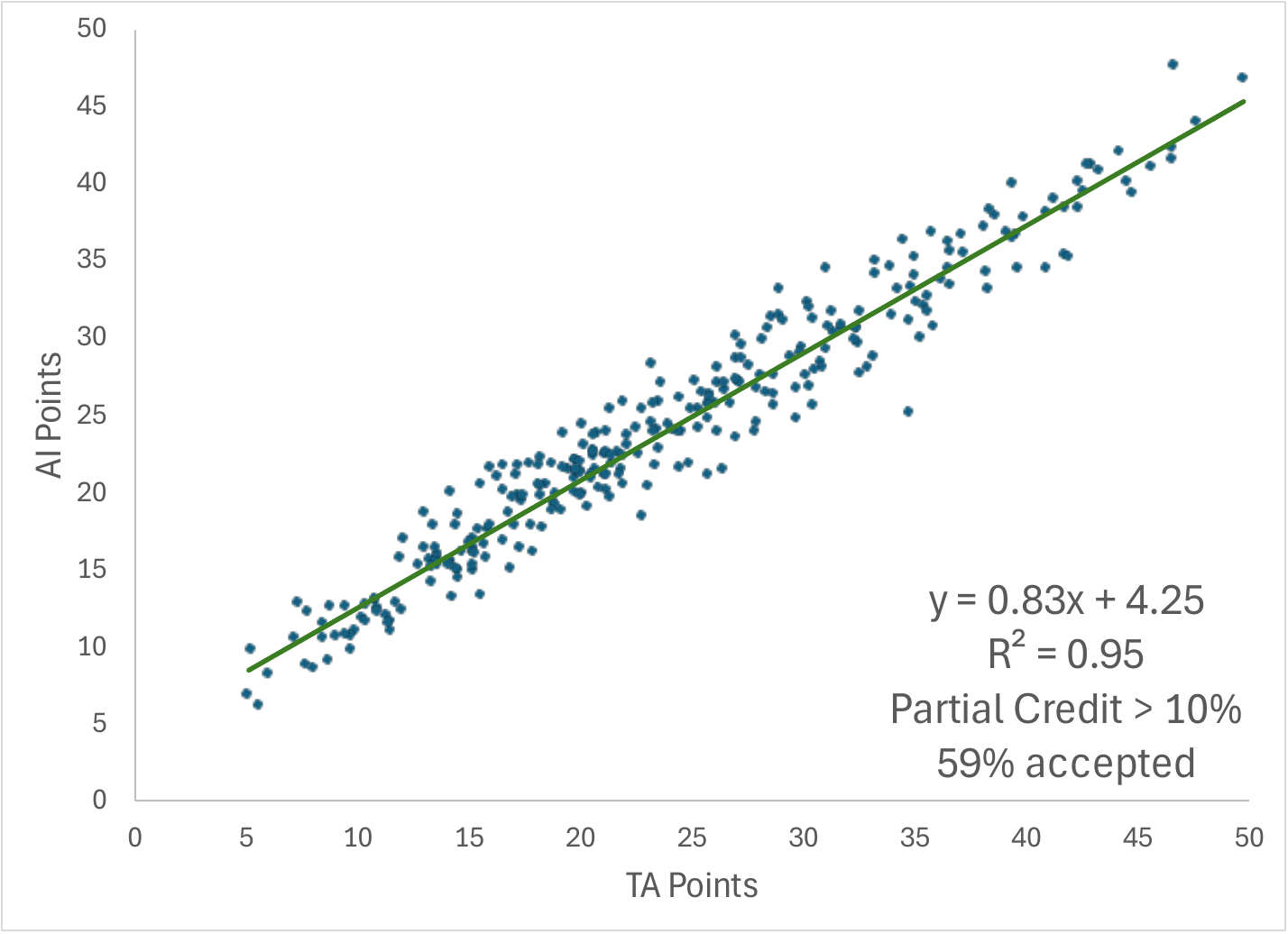}
  \includegraphics[width=0.33\textwidth]{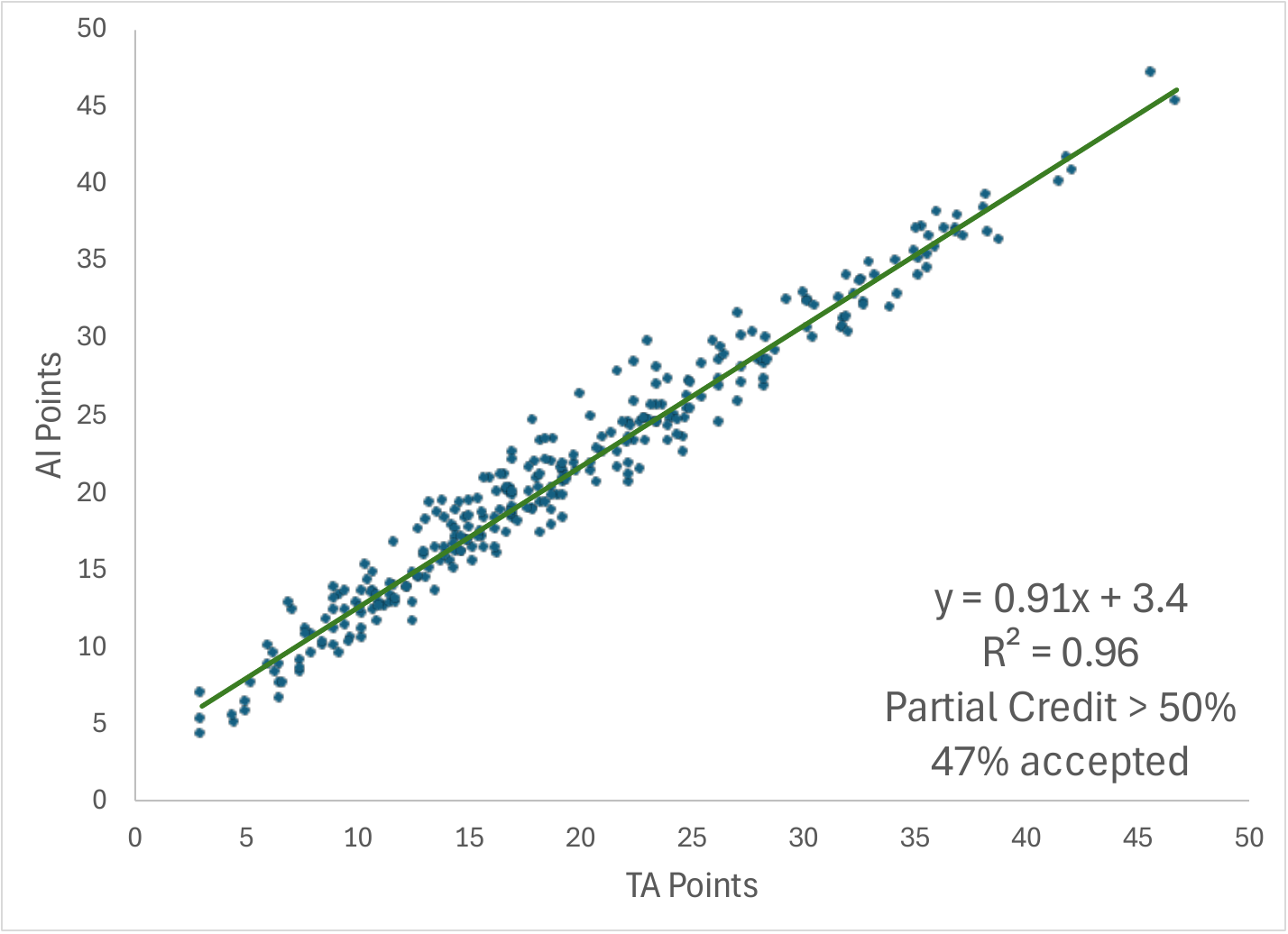}
  \includegraphics[width=0.33\textwidth]{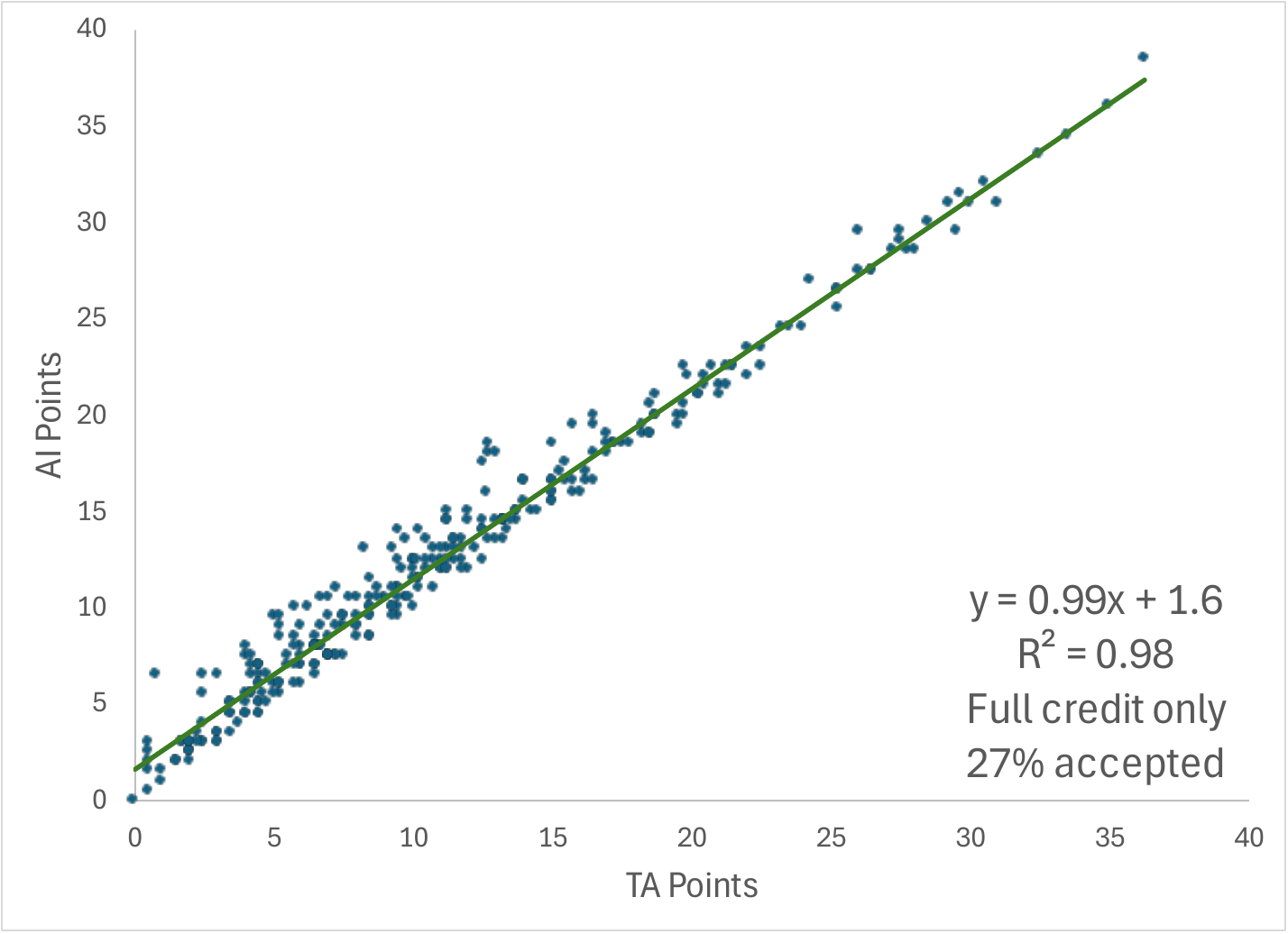}
 \caption{\ Scatterplots and linear regressions of the AI-assigned versus the TA-assigned points for the full exam. The panels, show the results when only accepting problem parts for which the AI assigned at least $10\%$ partial credit, at least half credit, and full credit, respectively.}
 \label{fig:partial}
\end{figure*}

The linear regressions of total AI-points versus true points echo this pattern, see Fig.~\ref{fig:partial}. With a generous 10\% cutoff, the slope is shallower and the intercept higher, reflecting consistent under-scoring at low thresholds and some baseline offset, yet the model still explains about 95\% of the variance in true class totals. At 50\% partial credit, the slope moves closer to unity, intercept shrinks, and the coefficient of determination ($R^2$) rises modestly, meaning the AI's total tallies align more faithfully with human scores when we only count mid- to high-confidence judgments. Finally, when we accept only full-credit items, the fit is almost perfect (slope $\approx1$, intercept minimal, $R^2\approx0.98$), but we're covering less than a third of all parts --- so the automated grading chunk is small but highly reliable.

Table~\ref{tab:partialqf} shows the effect of a 50\%-partial-credit ``mostly correct'' filter on the multiclass quality metrics for different question types. The improvements notable in the linear regression are not reflected in these measures. Instead, long answers appear to be graded much worse than before, however, looking at the small decrease in number of samples, it turns out that most of these answers had been mostly correct to begin with.

\begin{table*}
\small\centering
  \caption{\ Quality measures for the AI grading results in order of decreasing F1, filtered for problem parts where the AI-assigned partial credit is at least 50\%.}
  \label{tab:partialqf}
  \begin{tabular*}{0.8\textwidth}{@{\extracolsep{\fill}}lrrrlllll}
    \hline
 Type & Parts & Samples & Classes & Accuracy & Precision & Recall & F1 & normed F1\\
 \hline
Short Answer & 6 & 1089 & 9 & $0.76$ & $0.62$ & $0.57$ & $0.58$ & $0.52$ \\
Reaction & 3 & 502 & 6 & $0.7$ & $0.56$ & $0.55$ & $0.55$ & $0.46$ \\
Numerical & 18 & 1819 & 10 & $0.66$ & $0.39$ & $0.42$ & $0.39$ & $0.33$ \\
Multiple Choice & 9 & 1439 & 17 & $0.57$ & $0.33$ & $0.37$ & $0.33$ & $0.29$ \\
Symbolic & 3 & 507 & 6 & $0.78$ & $0.4$ & $0.54$ & $0.4$ & $0.28$ \\
\textit{All} & 46 & 6369 & 24 & $0.65$ & $0.3$ & $0.3$ & $0.29$ & $0.26$ \\
Long Answer & 1 & 230 & 5 & $0.82$ & $0.39$ & $0.42$ & $0.4$ & $0.25$ \\
\textit{Combination} & 4 & 595 & 13 & $0.41$ & $0.19$ & $0.24$ & $0.2$ & $0.14$ \\
Drawing & 1 & 163 & 5 & $0.61$ & $0.21$ & $0.28$ & $0.21$ & $0.01$ \\
Graphing & 1 & 25 & 4 & $0.6$ & $0.21$ & $0.18$ & $0.19$ & $-0.077$ \\
    \hline
  \end{tabular*}
\end{table*}

A lower partial-credit threshold gives broad coverage at the cost of extra checking on borderline items. If you prioritize near-perfect agreement, demand full credit and manually review the rest. Most instructors will likely choose an intermediate threshold, say 50\%, where the AI handles roughly half the work with strong regressions and F1, near its peak, leaving human graders free to focus on the trickiest, most ambiguous student responses. This layered workflow harnesses AI speed without sacrificing score fidelity on critical items.


\subsection{Filtering based on Bayesian statistics}
\begin{figure*}
\centering
  \includegraphics[width=0.94\textwidth]{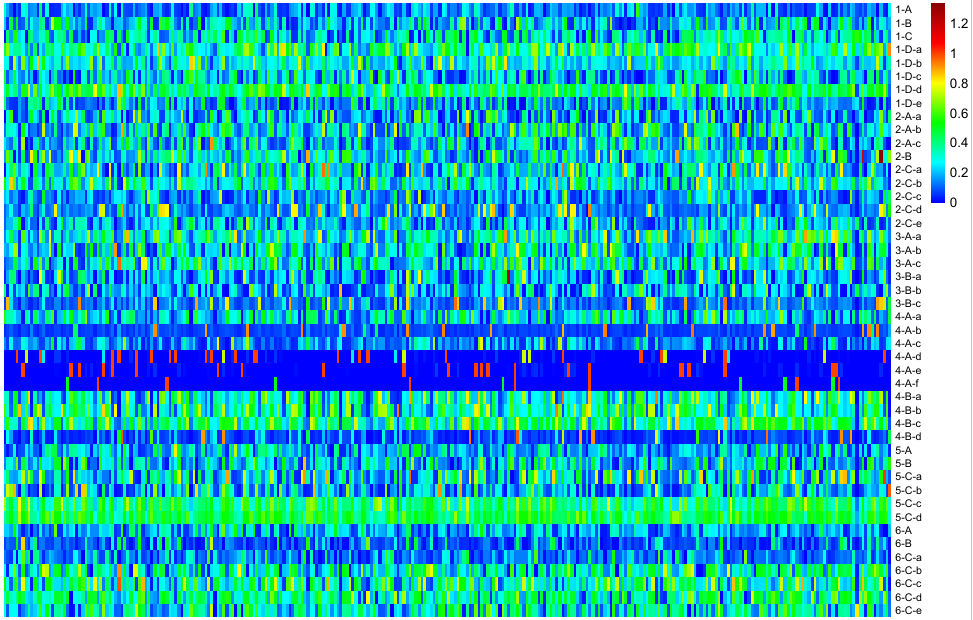}

  \caption{\ The matrix $\text{Risk}_{ij}$; differences are indicated by colors from blue (perfect agreement) to red (opposite of expectations). The dark red indicates the cases in which the absolute difference between the AI-assigned and the expected score was bigger than one, which can happen if the AI assigned more than the maximum points available or if it assigned negative points (see~Sect.~\ref{sec:fn}); the former kind of overflow happened twice for problem part 2-B and once for problem part 3-B-a (both numerical problems), the latter once for problem part 5-C-a (a multiple-choice problem)).}
  \label{fig:risk}
\end{figure*}

Figure~\ref{fig:risk} illustrates the difference between AI-grading and expectations based on IRT, see Equation~\ref{eq:risk}; the rows represent problem parts, the columns the participating students. The blue lines indicate high agreement between AI-grading and expectations for a particular problem part, while green, yellow, orange, and red indicate unexpected results. For example, the mostly-blue row for problem part 6-C-a corresponds to a short-answer problem, while 6-C-b involves drawing.

By sweeping $r$ from 0 (only perfectly expected scores) up to $1$ (all scores accepted), instructors can tune a trade-off:
\begin{itemize}
  \item At low $r$, only AI judgments very close to the IRT expectation are trusted --- yielding high confidence but low coverage.  
  \item As $r$ increases, more AI decisions fall within the tolerance, boosting automation at the cost of admitting more surprising (and potentially incorrect) scores.  
  \item Metrics like the coefficient of determination $R^2$ between accepted AI totals and true totals improve as $r$ tightens, while the fraction of parts requiring human review grows.  
\end{itemize}

\begin{figure}
 \centering
  \includegraphics[width=0.49\textwidth]{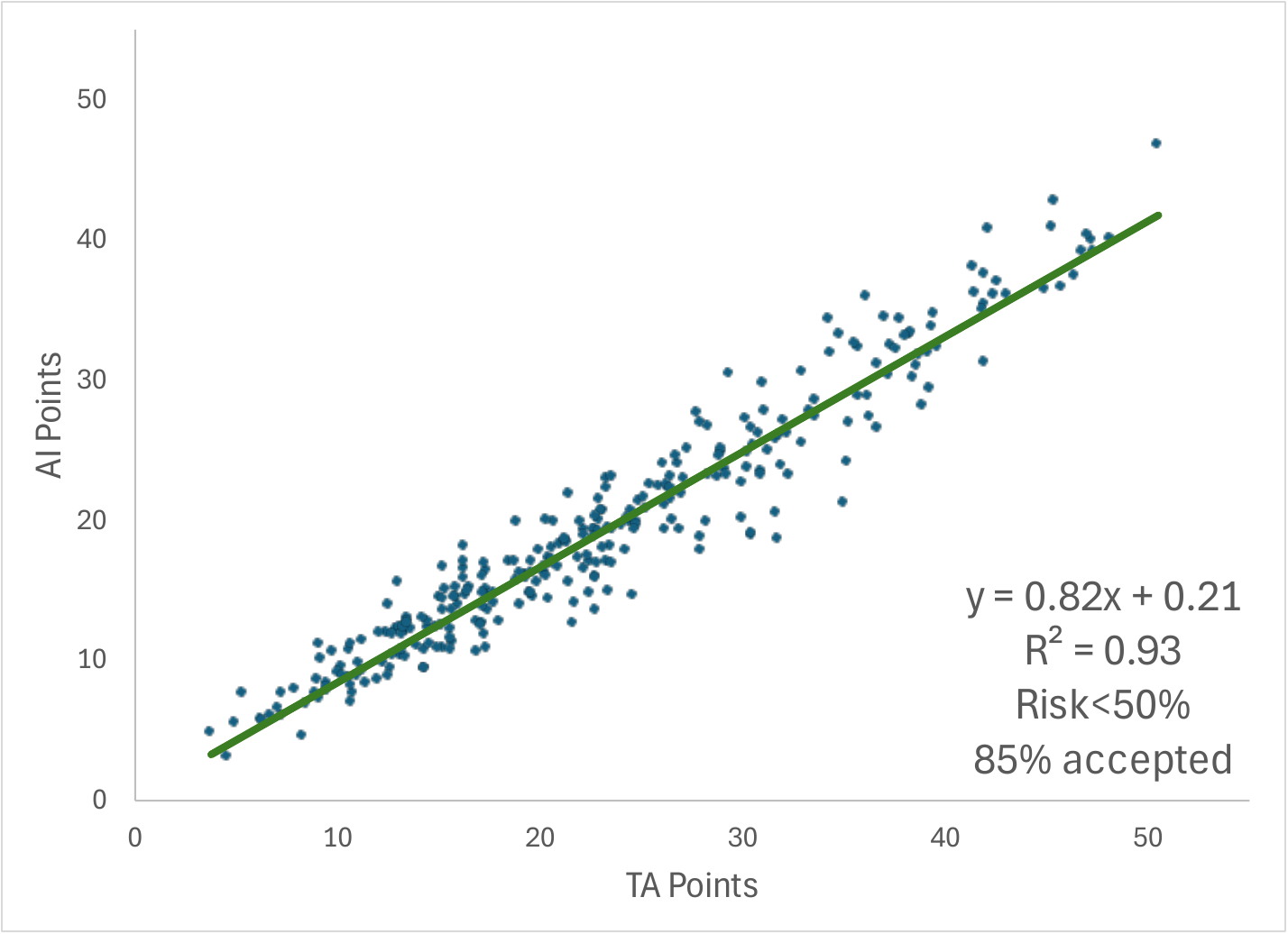}
 \caption{\ Scatterplot and linear regression of the AI-assigned versus the TA-assigned points for the full exam, filtered for problem parts where the AI-assigned partial credit deviates no more than 50\% from the expected partial credit.}
 \label{fig:riskreg}
\end{figure}

We chose $r=0.5$, which means that the AI's judgement can be up to $50\%$ away from the expected value, which results in an $85\%$ acceptance; Figure~\ref{fig:riskreg} shows the result. The linear regression shows clear improvement over the raw, unfiltered results.

Table~\ref{tab:riskqf} shows the quality measures for the risk-based filtering; these are only marginally improved compared to  the unfiltered results in Table~\ref{tab:allqf}. Once again, the same question types emerge as the most reliably graded.

\begin{table*}
\small\centering
  \caption{\ Multiclass quality measures for the AI grading results in order of decreasing F1, filtered for problem parts where the AI-assigned partial credit deviates no more than 50\% from the expected partial credit.}
  \label{tab:riskqf}
  \begin{tabular*}{0.8\textwidth}{@{\extracolsep{\fill}}lrrrlllll}
    \hline
 Type & Parts & Samples & Classes& Accuracy & Precision & Recall & F1 &normed F1\\
 \hline
Long Answer & 1 & 274 & 5 & $0.81$ & $0.71$ & $0.68$ & $0.68$ & $0.59$ \\
Short Answer & 6 & 1474 & 9 & $0.73$ & $0.65$ & $0.6$ & $0.6$ & $0.55$ \\
Reaction & 3 & 749 & 6 & $0.7$ & $0.66$ & $0.61$ & $0.63$ & $0.55$ \\
Multiple Choice & 9 & 2081 & 18 & $0.7$ & $0.4$ & $0.43$ & $0.41$ & $0.37$ \\
Numerical & 18 & 4750 & 9 & $0.64$ & $0.47$ & $0.43$ & $0.43$ & $0.36$ \\
Symbolic & 3 & 744 & 6 & $0.69$ & $0.49$ & $0.44$ & $0.44$ & $0.33$ \\
\textit{All} & 46 & 11599 & 24 & $0.64$ & $0.35$ & $0.33$ & $0.33$ & $0.3$ \\
Drawing & 1 & 230 & 5 & $0.58$ & $0.34$ & $0.38$ & $0.33$ & $0.16$ \\
\textit{Combination} & 4 & 1019 & 13 & $0.35$ & $0.21$ & $0.22$ & $0.2$ & $0.13$ \\
Graphing & 1 & 278 & 5 & $0.3$ & $0.061$ & $0.2$ & $0.094$ & $-0.13$ \\
\hline
  \end{tabular*}
\end{table*}


\subsubsection{Filtering by problem type}
\begin{figure*}
 \centering
  \includegraphics[width=0.33\textwidth]{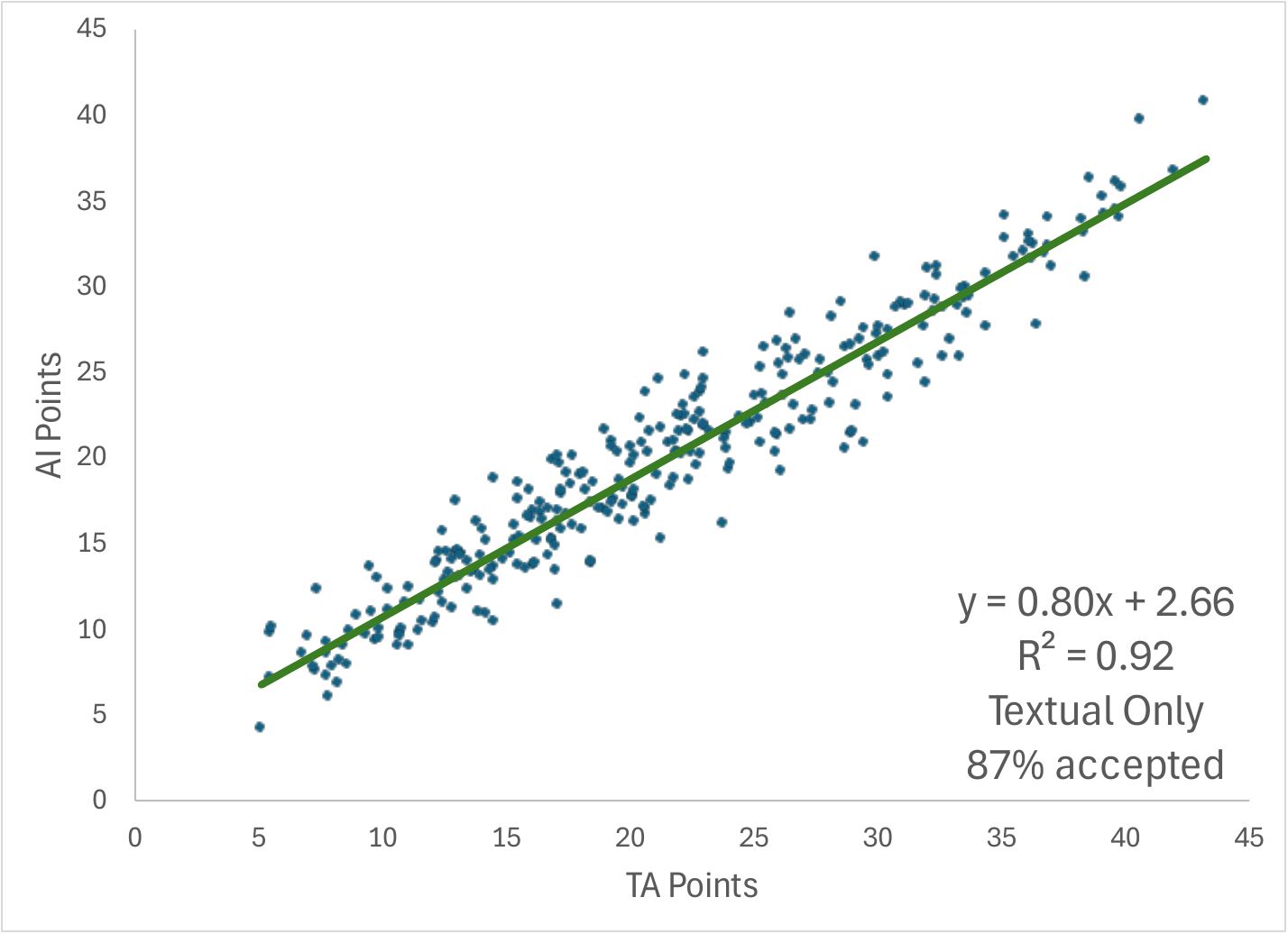}
  \includegraphics[width=0.33\textwidth]{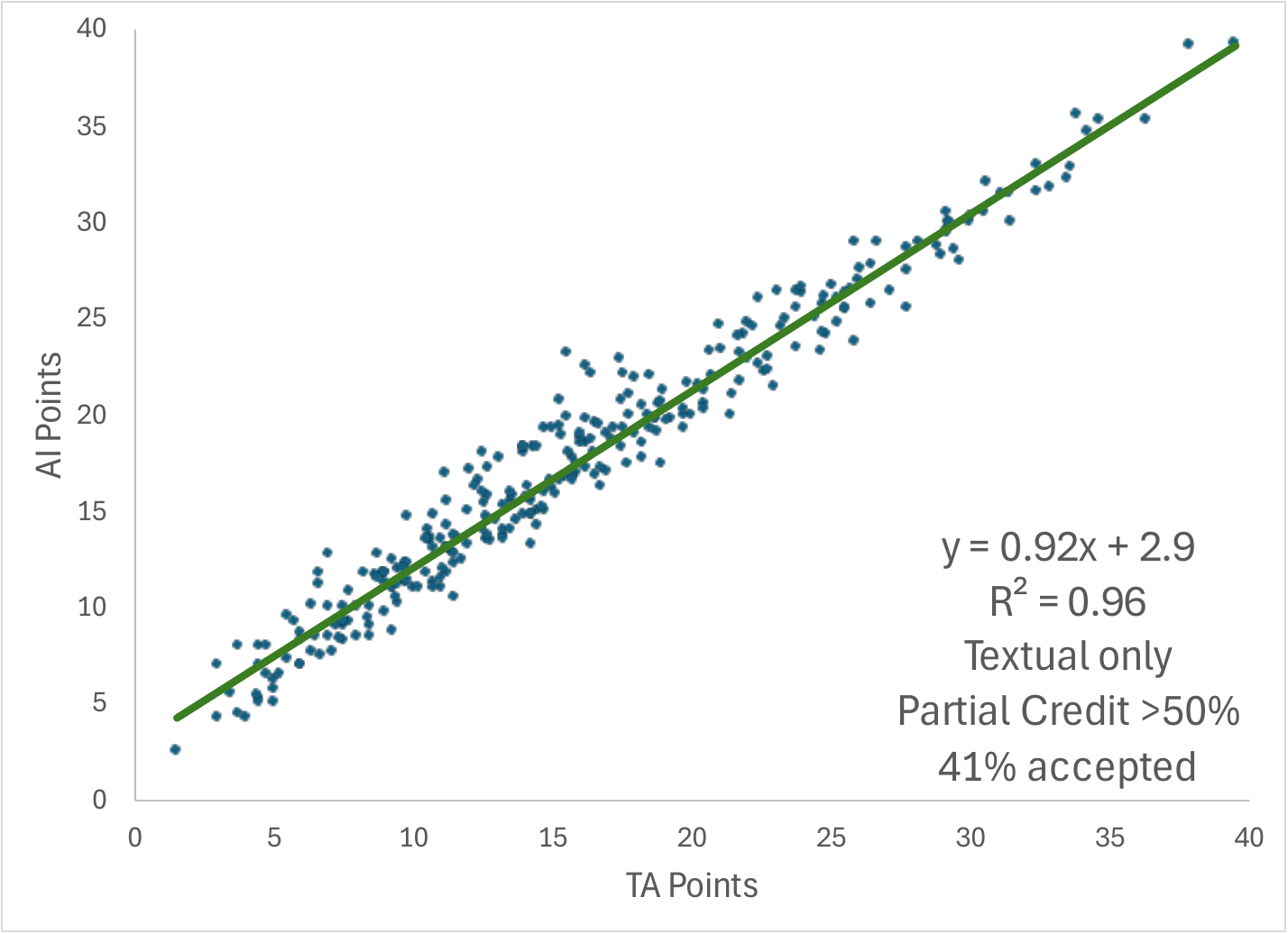}
  \includegraphics[width=0.33\textwidth]{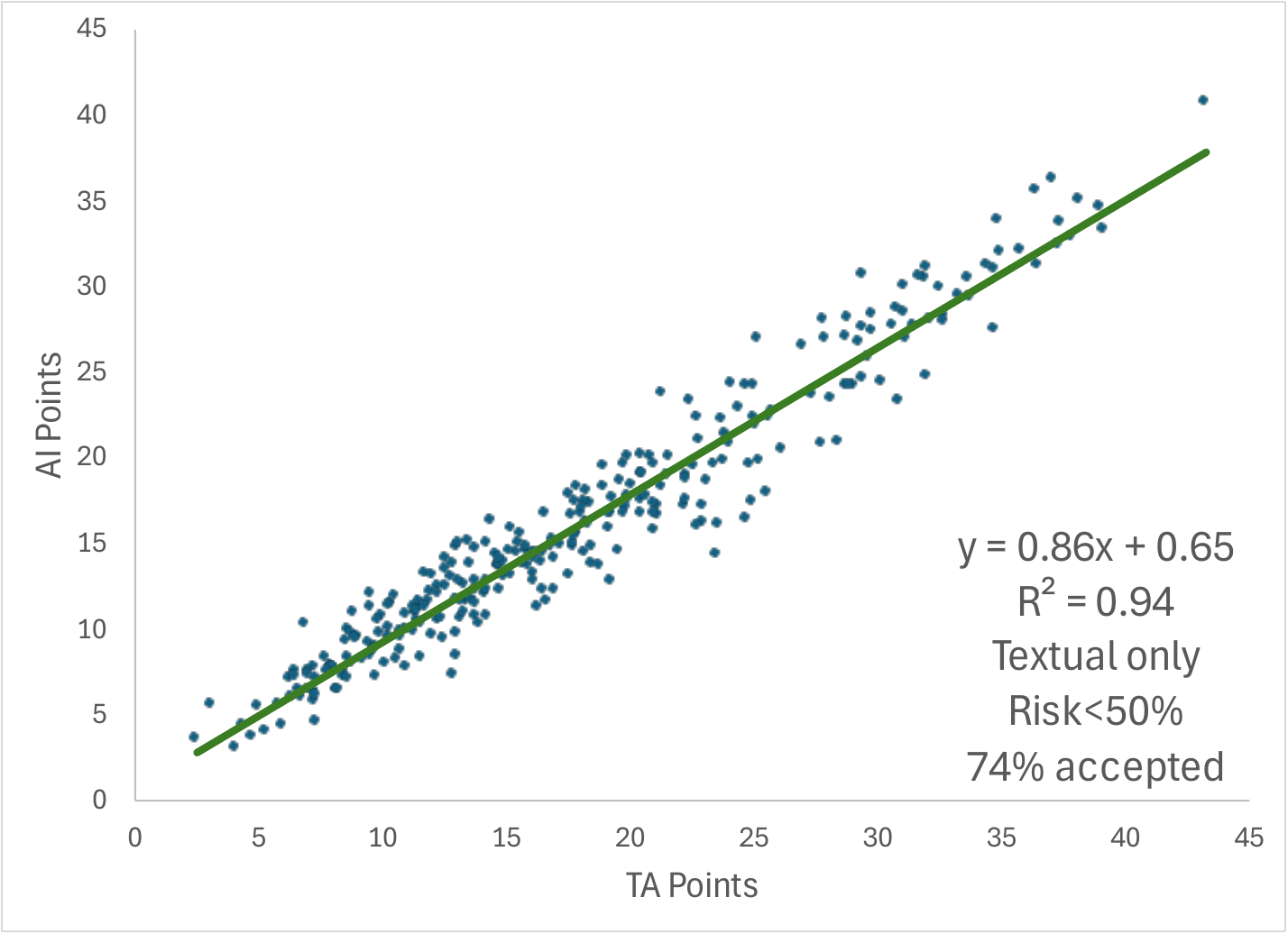}
 \caption{\ Scatterplots and linear regressions of the AI-assigned versus the TA-assigned points for only textual problems. The panels show the results when applying only this filter, when in addition only accepting problem parts for which the AI assigned at least $50\%$ partial credit, and when in addition only accepting less than 50\% risk based on Bayesian statistics.}
 \label{fig:textual}
\end{figure*}

Probably the most straightforward way of filtering the grading decisions is to a priory leave the grading of problem parts that include drawing or graphing to humans. Figure~\ref{fig:textual} shows the results when applying only this filter, when in addition only accepting problem parts for which the AI assigned at least $50\%$ partial credit, and when in addition only accepting less than 50\% risk based on Bayesian statistics.

Adding filters on top of the problem-type selection leads to marginal improvements, where however the risk-based filter lowers the false positives (as evidenced by the lower offset) while only rejecting an additional 8\%. 
%
%
\section{Discussion}

The results of this study underscore both the potential and current limitations of employing artificial intelligence (AI) systems for grading and assessment in chemistry education. Consistent with prior research, we found that AI-based scoring aligns closely with human grading for textual answers and chemical reaction equations, indicating that AI can effectively manage routine assessment tasks, thus allowing instructors and teaching assistants more time for nuanced instructional activities~\cite{Fernandez2024,Floden2025}. However, significant discrepancies remain for question formats involving graphical representations, echoing earlier findings that even advanced multimodal AI models struggle with diagrammatic reasoning and chemical structure interpretation~\cite{Ryan2019,kortemeyer2025multilingual}.

Our page-by-page rubric-based grading approach proved practical and intuitive, providing a clear method for instructors to integrate AI into existing grading workflows. This aligns with literature emphasizing intuitive and transparent grading practices to facilitate adoption among instructors~\cite{Azad2020}. Given these results, we recommend initially restricting AI-assisted grading primarily to textual responses and chemical equations, while reserving graphical and complex problem-solving tasks for human graders.

We evaluated several filtering strategies to optimize AI-human grading alignment:
\begin{itemize}
\item The partial-credit filter is intuitively appealing, as it benefits students by avoiding unwarranted score reductions. The main trade-off is increased human grading workload for partial-credit decisions, particularly as exam complexity grows.
\item The risk-based filter, employing Bayesian statistical methods, demonstrated greater efficiency in aligning scores and reducing TA workload. This approach aligns with recent studies advocating Bayesian and probabilistic methods in educational assessment to systematically manage grading uncertainties~\cite{kortemeyer2025assessing,Zitzmann2024}. However, its complexity and lower transparency may pose challenges for student acceptance.
\item The problem-type-based filter straightforwardly eliminates AI grading for historically challenging question formats, significantly improving reliability. Incorporating a risk-based filter on top provided minor additional improvements in grading alignment at a modest increase in human workload, supporting recommendations for targeted AI application in chemistry education~\cite{Ryan2019}.
\end{itemize}

The variability observed in AI and human grader agreement highlights the need for continued oversight, particularly in high-stakes assessments. The high $R^2$ values observed confirm that AI grading decisions are systematic and consistent with expert judgments, supporting literature indicating that AI can reliably mirror human grading patterns under controlled conditions~\cite{Floden2025,Emirtekin2025}. Our inter-run reliability analysis indicates that averaging several, independent grading runs could further improve the accuracy of the results. 

Implementing AI in formative and summative assessments also raises broader considerations about feedback quality and student perceptions. Future research must investigate student trust and perceived fairness, crucial elements influencing educational outcomes and acceptance of AI technologies. Recent studies suggest that clear explanations and transparency significantly enhance student acceptance of AI-graded assessments~\cite{Emirtekin2025}. Thus, ensuring transparency, accountability, and pedagogical effectiveness should remain central goals in AI-enhanced assessment frameworks~\cite{Azad2020,Fernandez2024}.

The purpose of our study was to explore the feasibility of offering meaningful, open-ended assessment more frequently to rising numbers of students. Thus, the goal is efficiency, not necessarily cost savings.
Nevertheless, it is instructive to look at cost: grading the 296 students by TAs cost $\$3,500/459\times296\approx\$2,250$. As only half of the results from the AI were deemed trustworthy, the cost using AI would be $\$100+\frac12\$2,250\approx\$1,225$. Not counted is the time required for depersonalization, as that would not be necessary outside of our IRB-protocol, and the time required for scanning (as this was done by senior personnel for our study, it cost \$1,400 in hourly-salary equivalent; it should be noted, though, that already today, many departments routinely scan their exams for evidence or for paperless grading). In both scenarios, students may appeal their scores, which requires additional faculty and staff time; this might happen more frequently with AI-assigned scores. Thus, AI grading would not necessarily be a huge cost-savings; instead, it would allow instructors to continue offering open-ended exam questions even in times of rising student numbers and stagnant resources.
%
%

\section{Limitations}\label{sec:limitations}
We did not perform parallel double scoring to estimate inter-rater reliability. Implementing a second, fully independent scoring stream would have required suppressing the calibration and adjudication steps described above and preventing the instructor from recalling prior decisions, thereby deviating from the authentic grading workflow.  Consequently, our labels represent the as-implemented institutional scoring rather than an independently replicated consensus. Residual rater error is possible; however, the design (single rater per item, upfront co-grading with a detailed rubric, documented adjudications of edge cases, instructor audit, and a post-exam appeals process) was intended to minimize idiosyncratic variance. Future work could assess inter-rater reliability on a stratified subsample or via a generalizability analysis, but this was outside the scope of the present, practice-grounded study.

The results presented here are specific to our exam, the model used (GPT-o4-mini), and the prompting; while we have found similar results with GPT-4o grading a thermodynamics exam~\cite{kortemeyer2025assessing} and GPT-5 grading a calculus exam~\cite{kortemeyer2025artificial}, results are not necessarily generalizable across models and prompts. For example, the negative points assigned to multiple-choice problems discussed in Sect.~\ref{sec:fn} could likely have been avoided with a more carefully formulated rubric (``prompting''), however, post-hoc manipulation of the rubric might defeat the purpose of modeling an authentic scenario.

\section{Conclusions}

In this study, we explored the effectiveness of AI-assisted grading for open-ended, handwritten general chemistry exams, aiming to balance efficiency with grading quality in large-enrollment courses. By comparing AI-generated scores to expert human graders, we demonstrated that carefully calibrated AI systems can substantially reduce grading workloads without sacrificing accuracy. Grading accuracy could be further improved by averaging multiple, independent runs. Our analysis indicated that deploying confidence filtering yields more dependable identification of fully correct answers, significantly streamlining grading when used selectively.

Based on these findings, we recommend that instructors initially deploy AI grading for low-stakes assessments such as quizzes and homework to establish confidence in the system. For high-stakes exams, a hybrid approach is most effective: instructors can confidently accept AI-graded full-credit responses, reserving human grading primarily for partially correct or ambiguous answers. Furthermore, utilizing AI-generated grading as a first-pass strategy allows teaching assistants to focus their expertise on nuanced student misconceptions and deeper feedback.

Pragmatically, instructors should clearly communicate to students the role of AI in grading processes, emphasizing transparency and fairness. Continuous monitoring and periodic recalibration of AI models are essential to maintain reliability. Ultimately, integrating AI-assisted grading thoughtfully into chemistry courses not only alleviates grading burdens but also enhances overall educational quality by enabling instructors to dedicate more time to meaningful interactions and instructional activities.

\section*{Author contributions}
\textbf{Jan Cvengros:} conceptualization (equal), data curation (lead), investigation (equal), resources (lead), supervision (equal), writing -- review and editing (equal). \textbf{Gerd Kortemeyer:} conceptualization (equal), formal analysis (lead), investigation (equal), methodology (lead), software (lead), supervision (equal), validation (lead), visualization (lead), writing -- original draft preparation (lead), writing -- review and editing (equal).

\section*{Conflicts of interest}
There are no conflicts to declare.

\section*{Ethical Considerations}
This study was approved by the ETH Zurich Ethics Commission (IRB) as protocol EK 2023-N-286. Students gave voluntary informed consent for participation.

\section*{Data availability}
The code for grading exams can be found at \url{https://gitlab.ethz.ch/ethel/etheltest/-/tree/main/ChemExamProject}.

\section*{Acknowledgements}

We would like to thank the students who chose to participate in our study. We would also like to thank Anna Kortemeyer for her help with data entry. We would also like to thank the reviewers of this journal for their insightful and constructive criticism.



\balance


\bibliography{chemexam} 

@article{kortemeyer2025artificial,
  title={Artificial-Intelligence Grading Assistance for Handwritten Components of a Calculus Exam},
  author={Kortemeyer, Gerd and Caspar, Alexander and Horica, Daria},
  journal={arXiv preprint arXiv:2510.05162},
  year={2025}
}

@article{impey2025using,
  title={Using large language models for automated grading of student writing about science},
  author={Impey, Chris and Wenger, Matthew and Garuda, Nikhil and Golchin, Shahriar and Stamer, Sarah},
  journal={International Journal of Artificial Intelligence in Education},
  pages={1--35},
  year={2025},
  publisher={Springer}
}

@inproceedings{zhu2025_ijcai,
  title={Enhancing Automated Grading in Science Education through LLM-Driven Causal Reasoning and Multimodal Analysis},
  author={Zhu, Haohao and Li, Tingting and He, Peng and Zhou, Jiayu},
  year={2025},
  booktitle={Proceedings of the Thirty-Fourth International Joint Conference on Artificial Intelligence},
  organization={International Joint Conferences on Artificial Intelligence Organization}
}

@article{chu2025_edm_graderag,
  title={Enhancing LLM-Based Short Answer Grading with Retrieval-Augmented Generation},
  author={Chu, Yucheng and He, Peng and Li, Hang and Han, Haoyu and Yang, Kaiqi and Xue, Yu and Li, Tingting and Krajcik, Joseph and Tang, Jiliang},
  journal={arXiv preprint arXiv:2504.05276},
  year={2025}
}

@inproceedings{asag2024,
  title={Asag2024: A combined benchmark for short answer grading},
  author={Meyer, G{\'e}r{\^o}me and Breuer, Philip and F{\"u}rst, Jonathan},
  booktitle={Proceedings of the 2024 on ACM Virtual Global Computing Education Conference V. 2},
  pages={322--323},
  year={2024}
}

@inproceedings{cohn2024_cot,
  title={A chain-of-thought prompting approach with llms for evaluating studentsÕ formative assessment responses in science},
  author={Cohn, Clayton and Hutchins, Nicole and Le, Tuan and Biswas, Gautam},
  booktitle={Proceedings of the AAAI conference on artificial intelligence},
  volume={38},
  number={21},
  pages={23182--23190},
  year={2024}
}

@article{grevisse2024_bmcmeded,
  title={LLM-based automatic short answer grading in undergraduate medical education},
  author={Gr{\'e}visse, Christian},
  journal={BMC Medical Education},
  volume={24},
  number={1},
  pages={1060},
  year={2024},
  publisher={Springer}
}

@article{latif2024_patterns,
  title={Fine-tuning ChatGPT for automatic scoring},
  author={Latif, Ehsan and Zhai, Xiaoming},
  journal={Computers and Education: Artificial Intelligence},
  volume={6},
  pages={100210},
  year={2024},
  publisher={Elsevier}
}

@inproceedings{sonkar2024automated,
  title={Automated long answer grading with RiceChem dataset},
  author={Sonkar, Shashank and Ni, Kangqi and Tran Lu, Lesa and Kincaid, Kristi and Hutchinson, John S and Baraniuk, Richard G},
  booktitle={International Conference on Artificial Intelligence in Education},
  pages={163--176},
  year={2024},
  organization={Springer}
}

@article{lee2024_patterns,
  title={Applying large language models and chain-of-thought for automatic scoring},
  author={Lee, Gyeong-Geon and Latif, Ehsan and Wu, Xuansheng and Liu, Ninghao and Zhai, Xiaoming},
  journal={Computers and Education: Artificial Intelligence},
  volume={6},
  pages={100213},
  year={2024},
  publisher={Elsevier}
}

@article{plyer2024_softgrading,
  title={Implementation of a soft grading system for chemistry in a Moodle plugin: reaction handling},
  author={Plyer, Louis and Marcou, Gilles and Perves, C{\'e}line and Bonachera, Fanny and Varnek, Alexander},
  journal={Journal of Cheminformatics},
  volume={16},
  number={1},
  pages={90},
  year={2024},
  publisher={Springer}
}

@inproceedings{yamtinah2024_icce,
  title={Leveraging Generative AI for Automatic Scoring in Chemistry Education: A Web Based Approach to Assessing Conceptual Understanding of Colligative Properties},
  author={Yamtinah, Sri and Ramadhani, Dimas Gilang and Wiyarsi, Antuni and Widarti, Hayuni Retno and Shidiq, Ari Syahidul},
  booktitle={International Conference on Computers in Education},
  year={2024}
}

@inproceedings{li2023learning,
  title={Learning when to defer to humans for short answer grading},
  author={Li, Zhaohui and Zhang, Chengning and Jin, Yumi and Cang, Xuesong and Puntambekar, Sadhana and Passonneau, Rebecca J},
  booktitle={International Conference on Artificial Intelligence in Education},
  pages={414--425},
  year={2023},
  organization={Springer}
}

@article{kane2013validating,
  title={Validating the interpretations and uses of test scores},
  author={Kane, Michael T},
  journal={Journal of educational measurement},
  volume={50},
  number={1},
  pages={1--73},
  year={2013},
  publisher={Wiley Online Library}
}

@article{brennan1992generalizability,
  title={Generalizability theory},
  author={Brennan, Robert L},
  journal={Educational Measurement: Issues and Practice},
  volume={11},
  number={4},
  pages={27--34},
  year={1992},
  publisher={Wiley Online Library}
}

@article{lawrence1989concordance,
  title={A concordance correlation coefficient to evaluate reproducibility},
  author={Lawrence, I and Lin, Kuei},
  journal={Biometrics},
  pages={255--268},
  year={1989},
  publisher={JSTOR}
}

@article{kendall1939problem,
  title={The problem of m rankings},
  author={Kendall, Maurice G and Smith, B Babington},
  journal={The annals of mathematical statistics},
  volume={10},
  number={3},
  pages={275--287},
  year={1939},
  publisher={JSTOR}
}

@article{shrout1979intraclass,
  title={Intraclass correlations: uses in assessing rater reliability.},
  author={Shrout, Patrick E and Fleiss, Joseph L},
  journal={Psychological bulletin},
  volume={86},
  number={2},
  pages={420},
  year={1979},
  publisher={American Psychological Association}
}

@article{gomez2025student,
  title={Student success and experience in a flipped, senior physical chemistry course spanning before and after the COVID-19 pandemic},
  author={Gomez, Trisha M and Luciano, Charmaine and Nguyen, Tam and Villafa{\~n}e, Sachel M and Groves, Michael N},
  journal={Chemistry Education Research and Practice},
  volume={26},
  number={1},
  pages={210--230},
  year={2025},
  publisher={Royal Society of Chemistry}
}

@article{baade2025seeing,
  title={ÔSeeingÕchemistry: investigating the contribution of mental imagery strength on studentsÕ thinking in relation to visuospatial problem solving in chemistry},
  author={Baade, Lauren and Kartsonaki, Effie and Khosravi, Hassan and Lawrie, Gwendolyn A},
  journal={Chemistry Education Research and Practice},
  volume={26},
  number={1},
  pages={65--87},
  year={2025},
  publisher={Royal Society of Chemistry}
}

@article{stammes2025drawing,
  title={Drawing meaning from student-generated drawings: exploring chemistry teachersÕ noticing},
  author={Stammes, Hanna and de Putter-Smits, Lesley},
  journal={Chemistry Education Research and Practice},
  volume={26},
  number={2},
  pages={494--507},
  year={2025},
  publisher={Royal Society of Chemistry}
}

@article{ye2025just,
  title={`It just feels like it's gonna be so very long?' Exploring the resources used by university students in noticing, navigating, and resolving issues during math-intensive problem solving in chemistry},
  author={Ye, Sofie and Jacobsson, Magnus and Elmgren, Maja and Ho, Felix M},
  journal={Chemistry Education Research and Practice},
  volume={26},
  number={2},
  pages={377--397},
  year={2025},
  publisher={Royal Society of Chemistry}
}

@article{grieger2025utility,
  title={Utility of creative exercises as an assessment tool for revealing student conceptions in organic chemistry},
  author={Grieger, Krystal and Leontyev, Alexey},
  journal={Chemistry Education Research and Practice},
  year={2025},
  pages={603--618},
  publisher={Royal Society of Chemistry}
}

@article{pentucci2025developing,
  title={Developing self-reflection in students: a case study in chemistry education},
  author={Pentucci, Maila and Mascitti, Andrea and d'Alessandro, Nicola and Tonucci, Lucia and Coccia, Francesca},
  journal={Chemistry Education Research and Practice},
  year={2025},
  pages={adv.art.},
  publisher={Royal Society of Chemistry}
}

@article{towns2014guide,
  title={Guide to developing high-quality, reliable, and valid multiple-choice assessments},
  author={Towns, Marcy H},
  journal={Journal of Chemical Education},
  volume={91},
  number={9},
  pages={1426--1431},
  year={2014},
  publisher={ACS Publications}
}

@article{mullen2012short,
  title={Short answer versus multiple choice examination questions for first year chemistry},
  author={Mullen, Kathleen and Schultz, Madeleine},
  journal={International Journal of Innovation in Science and Mathematics Education},
  volume={20},
  number={3},
  pages={1--18},
  year={2012}
}

@article{clark2020testing,
  title={Testing in the time of {COVID}-19: A sudden transition to unproctored online exams},
  author={Clark, Ted M and Callam, Christopher S and Paul, Noel M and Stoltzfus, Matthew W and Turner, Daniel},
  journal={Journal of chemical education},
  volume={97},
  number={9},
  pages={3413--3417},
  year={2020},
  publisher={ACS Publications}
}

@article{stowe2019assessment,
  title={Assessment in chemistry education},
  author={Stowe, Ryan L and Cooper, Melanie M},
  journal={Israel Journal of chemistry},
  volume={59},
  number={6-7},
  pages={598--607},
  year={2019},
  publisher={Wiley Online Library}
}

@article{gardner2023challenges,
  title={The challenges and value of undergraduate oral exams in the physical chemistry classroom: A useful tool in the assessment toolbox},
  author={Gardner, David E and Giordano, Andrea N},
  journal={Journal of chemical education},
  volume={100},
  number={5},
  pages={1705--1709},
  year={2023},
  publisher={ACS Publications}
}

@article{zoller2002algorithmic,
  title={Algorithmic, {LOCS} and {HOCS} (chemistry) exam questions: Performance and attitudes of college students},
  author={Zoller, Uri},
  journal={International Journal of Science Education},
  volume={24},
  number={2},
  pages={185--203},
  year={2002},
  publisher={Taylor \& Francis}
}

@incollection{offerdahl2019formative,
  title={Formative assessment to improve student learning in biochemistry},
  author={Offerdahl, Erika G and Arneson, Jessie B},
  booktitle={Biochemistry education: From theory to practice},
  pages={197--218},
  year={2019},
  publisher={ACS Publications}
}

@article{cho2010learning,
  author = {Cho, Young Hoan and Cho, Kwangsu},
  title = {Peer reviewers learn from giving comments},
  journal = {Instructional Science},
  volume = {39},
  number = {5},
  pages = {629--643},
  year = {2011},
  doi = {10.1007/s11251-010-9146-1}
}

@article{alsalmani23a,
  author = {AlSalmani, Ahmed and Babiker, Abdel and Abdallah, Rami},
  title = {Human-in-the-loop Approaches in Educational Assessment Systems: A Review},
  journal = {Educational Technology Research and Development},
  year = {2023},
  volume = {71},
  number = {2},
  pages = {279--299},
  doi = {10.1007/s11423-022-10180-8}
}

@article{Ryan2019,
  author = {Ryan, Stephanie A. C. and Stieff, Mike},
  title = {Drawing for Assessing Learning Outcomes in Chemistry},
  journal = {Journal of Chemical Education},
  volume = {96},
  number = {9},
  pages = {1813--1820},
  year = {2019},
  doi = {10.1021/acs.jchemed.9b00361}
}

@article{Zitzmann2024,
  author = {Zitzmann, Steffen and Orona, Gabe A. and Lohmann, Julian F. and K{\"o}nig, Claudia and Bardach, Lisa and Hecht, Markus},
  title = {Novick meets {Bayes}: Improving the assessment of individual students in educational practice and research by capitalizing on assessors' prior beliefs},
  journal = {Educational and Psychological Measurement},
  year = {2024},
  pages={483-506},
  doi = {10.1177/00131644241296139}
}

@article{Floden2025,
  author    = {Jonas Flod{\'e}n},
  title     = {Grading exams using large language models: A comparison between human and AI grading of exams in higher education using ChatGPT},
  journal   = {British Educational Research Journal},
  year      = {2025},
  volume    = {51},
  number    = {1},
  pages     = {201--224},
  doi       = {10.1002/berj.4069}
}

@article{Fernandez2024,
  author    = {Alberto A. Fern{\'a}ndez and Margarita L{\'o}pez-Torres and Jes{\'u}s J. Fern{\'a}ndez and Digna V{\'a}zquez-Garc{\'i}a},
  title     = {ChatGPT as an InstructorÕs Assistant for Generating and Scoring Exams},
  journal   = {Journal of Chemical Education},
  year      = {2024},
  volume    = {101},
  number    = {9},
  pages     = {3780--3788},
  doi       = {10.1021/acs.jchemed.4c00231}
}

@inproceedings{Azad2020,
  author    = {Sushmita Azad and Binglin Chen and Maxwell Fowler and Matthew West and Craig Zilles},
  title     = {Strategies for Deploying Unreliable {AI} Graders in High-Transparency High-Stakes Exams},
  booktitle = {Proceedings of the 21st International Conference on Artificial Intelligence in Education (AIED~2020)},
  series    = {Lecture Notes in Computer Science},
  volume    = {12163},
  pages     = {16--28},
  publisher = {Springer},
  address   = {Cham},
  year      = {2020},
  doi       = {10.1007/978-3-030-52237-7\_2}
}

@article{Emirtekin2025,
  author    = {Emrah Emirtekin},
  title     = {Large Language Model-Powered Automated Assessment: A Systematic Review},
  journal   = {Applied Sciences},
  year      = {2025},
  volume    = {15},
  number    = {10},
  pages     = {5683},
  doi       = {10.3390/app15105683}
}

@misc{euactobl,
  title = "{EU Artificial Intelligence Act, Article 26: Obligations of Deployers of High-Risk AI Systems}",
  howpublished = {\url{https://artificialintelligenceact.eu/article/26/}},
  month = jul,
  year = "2024",
  author = "{European Union}"
}

@article{hart1912general,
  title={General ability, its existence and nature},
  author={Hart, Bernard and Spearman, Charles},
  journal={British Journal of Psychology},
  volume={5},
  number={4},
  pages={51},
  year={1912},
  publisher={Cambridge University Press}
}

@article{thurstone1934vectors,
  title={The vectors of mind.},
  author={Thurstone, Louis Leon},
  journal={Psychological review},
  volume={41},
  number={1},
  pages={1},
  year={1934},
  publisher={Psychological Review Company}
}

@inproceedings{de2020case,
  title={A case for humans-in-the-loop: Decisions in the presence of erroneous algorithmic scores},
  author={De-Arteaga, Maria and Fogliato, Riccardo and Chouldechova, Alexandra},
  booktitle={Proceedings of the 2020 CHI Conference on Human Factors in Computing Systems},
  pages={1--12},
  year={2020}
}

@article{spearman1961general,
  title={``General Intelligence'' Objectively Determined and Measured.},
  journal={American Journal of Psychology},
  volume={15},
  pages={201-293},
  author={Spearman, Charles},
  year={1961}
}

@misc{euactannex,
  title = "{EU Artificial Intelligence Act, Annex III}",
  howpublished = {\url{https://artificialintelligenceact.eu/annex/3/}},
  month = jul,
  year = "2024",
  author = "{European Union}"
}

@inproceedings{li2023wrong,
  title={Am I wrong, or is the autograder wrong? Effects of AI grading mistakes on learning},
  author={Li, Tiffany Wenting and Hsu, Silas and Fowler, Max and Zhang, Zhilin and Zilles, Craig and Karahalios, Karrie},
  booktitle={Proceedings of the 2023 ACM Conference on International Computing Education Research-Volume 1},
  pages={159--176},
  year={2023}
}

@article{meyer2023chatgpt,
  title={{ChatGPT} and large language models in academia: opportunities and challenges},
  author={Meyer, Jesse G and Urbanowicz, Ryan J and Martin, Patrick CN and O'Connor, Karen and Li, Ruowang and Peng, Pei-Chen and Bright, Tiffani J and Tatonetti, Nicholas and Won, Kyoung Jae and Gonzalez-Hernandez, Graciela and others},
  journal={BioData Mining},
  volume={16},
  number={1},
  pages={20},
  year={2023},
  publisher={Springer}
}

@article{kasneci2023chatgpt,
  title={{ChatGPT} for good? On opportunities and challenges of large language models for education},
  author={Kasneci, Enkelejda and Se{\ss}ler, Kathrin and K{\"u}chemann, Stefan and Bannert, Maria and Dementieva, Daryna and Fischer, Frank and Gasser, Urs and Groh, Georg and G{\"u}nnemann, Stephan and H{\"u}llermeier, Eyke and others},
  journal={Learning and individual differences},
  volume={103},
  pages={102274},
  year={2023},
  publisher={Elsevier}
}

@article{yik2021development,
  title={Development of a machine learning-based tool to evaluate correct Lewis acid--base model use in written responses to open-ended formative assessment items},
  author={Yik, Brandon J and Dood, Amber J and De Arellano, Daniel Cruz-Ram{\'\i}rez and Fields, Kimberly B and Raker, Jeffrey R},
  journal={Chemistry Education Research and Practice},
  volume={22},
  number={4},
  pages={866--885},
  year={2021},
  publisher={Royal Society of Chemistry}
}

@article{martin2023machine,
  title={When a machine detects student reasoning: a review of machine learning-based formative assessment of mechanistic reasoning},
  author={Martin, Paul P and Graulich, Nicole},
  journal={Chemistry Education Research and Practice},
  volume={24},
  number={2},
  pages={407--427},
  year={2023},
  publisher={Royal Society of Chemistry}
}

@article{lawrie2023establishing,
  title={Establishing a delicate balance in the relationship between artificial intelligence and authentic assessment in student learning},
  author={Lawrie, Gwendolyn},
  journal={Chemistry Education Research and Practice},
  volume={24},
  number={2},
  pages={392--393},
  year={2023},
  publisher={Royal Society of Chemistry}
}

@article{kortemeyer2025multilingual,
  title={Multilingual performance of a multimodal artificial intelligence system on multisubject physics concept inventories},
  author={Kortemeyer, Gerd and Babayeva, Marina and Polverini, Giulia and Widenhorn, Ralf and Gregorcic, Bor},
  journal={Physical Review Physics Education Research},
  volume={21},
  number={2},
  pages={020101},
  year={2025},
  publisher={APS}
}

@article{kortemeyer2024grading,
  title={Grading assistance for a handwritten thermodynamics exam using artificial intelligence: An exploratory study},
  author={Kortemeyer, Gerd and N{\"o}hl, Julian and Onishchuk, Daria},
  journal={Physical Review Physics Education Research},
  volume={20},
  number={2},
  pages={020144},
  year={2024},
  publisher={APS}
}

@article{kortemeyer2025assessing,
  title={Assessing confidence in AI-assisted grading of physics exams through psychometrics: An exploratory study},
  author={Kortemeyer, Gerd and N{\"o}hl, Julian},
  journal={Physical Review Physics Education Research},
  volume={21},
  number={1},
  pages={010136},
  year={2025},
  publisher={APS}
}

@article{wan24,
  title = {Exploring generative {AI} assisted feedback writing for students' written responses to a physics conceptual question with prompt engineering and few-shot learning},
  author = {Wan, Tong and Chen, Zhongzhou},
  journal = {Phys. Rev. Phys. Educ. Res.},
  volume = {20},
  issue = {1},
  pages = {010152},
  numpages = {13},
  year = {2024},
  month = {Jun},
  publisher = {American Physical Society},
  doi = {10.1103/PhysRevPhysEducRes.20.010152},
  url = {https://link.aps.org/doi/10.1103/PhysRevPhysEducRes.20.010152}
}

@article{wilson22,
  title = {Classification of open-ended responses to a research-based assessment using natural language processing},
  author = {Wilson, Joseph and Pollard, Benjamin and Aiken, John M. and Caballero, Marcos D. and Lewandowski, H. J.},
  journal = {Phys. Rev. Phys. Educ. Res.},
  volume = {18},
  issue = {1},
  pages = {010141},
  numpages = {16},
  year = {2022},
  month = {Jun},
  publisher = {American Physical Society},
  doi = {10.1103/PhysRevPhysEducRes.18.010141},
  url = {https://link.aps.org/doi/10.1103/PhysRevPhysEducRes.18.010141}
}

@article{alsalmani23,
  title = {Assessing thinking skills in free-response exam problems: Pandemic online and in-person},
  author = {Al-Salmani, Fatema and Johnson, Jordan and Thacker, Beth},
  journal = {Phys. Rev. Phys. Educ. Res.},
  volume = {19},
  issue = {1},
  pages = {010131},
  numpages = {8},
  year = {2023},
  month = {May},
  publisher = {American Physical Society},
  doi = {10.1103/PhysRevPhysEducRes.19.010131},
  url = {https://link.aps.org/doi/10.1103/PhysRevPhysEducRes.19.010131}
}

@article{kortemeyer24aigrading,
  title = {Toward {AI} grading of student problem solutions in introductory physics: A feasibility study},
  author = {Kortemeyer, Gerd},
  journal = {Phys. Rev. Phys. Educ. Res.},
  volume = {19},
  issue = {2},
  pages = {020163},
  numpages = {20},
  year = {2023},
  month = {Nov},
  publisher = {American Physical Society},
  doi = {10.1103/PhysRevPhysEducRes.19.020163},
  url = {https://link.aps.org/doi/10.1103/PhysRevPhysEducRes.19.020163}
}

@misc{gpto4,
	author = {{OpenAI}},
	howpublished = {\url{https://platform.openai.com/docs/guides/reasoning}},
	title = {{Reasoning models}},
	year = {accessed July 2025}}

@article{kortemeyer2023performance,
 title={Performance of the pre-trained large language model {GPT-4} on automated short answer grading},
  author={Kortemeyer, Gerd},
  journal={Discover Artificial Intelligence},
  volume={4},
  number={1},
  pages={47},
  year={2024},
  publisher={Springer},
    address={New York, NY}
}

@article{ding2023students,
  title={Students' perceptions of using {ChatGPT} in a physics class as a virtual tutor},
  author={Ding, Lu and Li, Tong and Jiang, Shiyan and Gapud, Albert},
  journal={International Journal of Educational Technology in Higher Education},
  volume={20},
  number={1},
  pages={63},
  year={2023},
  publisher={Springer},
    address={New York, NY}
}

@article{polverini24,
	author = {Polverini, Giulia and Gregorcic, Bor},
	doi = {10.1103/PhysRevPhysEducRes.20.010109},
	issue = {1},
	journal = {Phys. Rev. Phys. Educ. Res.},
	month = {Feb},
	numpages = {16},
	pages = {010109},
	publisher = {American Physical Society},
	title = {Performance of {ChatGPT} on the test of understanding graphs in kinematics},
	url = {https://link.aps.org/doi/10.1103/PhysRevPhysEducRes.20.010109},
	volume = {20},
	year = {2024}}

@article{dahlkemper23,
	author = {Dahlkemper, Merten Nikolay and Lahme, Simon Zacharias and Klein, Pascal},
	doi = {10.1103/PhysRevPhysEducRes.19.010142},
	issue = {1},
	journal = {Phys. Rev. Phys. Educ. Res.},
	month = {Jun},
	numpages = {25},
	pages = {010142},
	publisher = {American Physical Society},
	title = {How do physics students evaluate artificial intelligence responses on comprehension questions? A study on the perceived scientific accuracy and linguistic quality of {ChatGPT}},
	url = {https://link.aps.org/doi/10.1103/PhysRevPhysEducRes.19.010142},
	volume = {19},
	year = {2023}}

@misc{azure,
	author = {{Microsoft}},
	howpublished = {\url{https://azure.microsoft.com/en-us/products/ai-services/}},
	title = {{Azure AI Services}},
	year = {accessed July 2025}}

@article{sinharay2006posterior,
	author = {Sinharay, Sandip and Johnson, Matthew S and Stern, Hal S},
	journal = {Applied Psychological Measurement},
	number = {4},
	pages = {298--321},
	publisher = {Sage Publications},
	title = {Posterior predictive assessment of item response theory models},
	volume = {30},
	year = {2006}}

@article{kortemeyer2019quick,
	author = {Kortemeyer, Gerd},
	journal = {The Physics Teacher},
	number = {9},
	pages = {608--610},
	publisher = {American Association of Physics Teachers},
	title = {Quick-and-Dirty Item Response Theory},
	volume = {57},
	year = {2019}}

@book{rasch,
	author = {Rasch, Georg},
	publisher = {ERIC},
	title = {Probabilistic models for some intelligence and attainment tests.},
	year = {1993}}

@article{pawl2013,
	author = {Pawl, Andrew and Teodorescu, Raluca and Peterson, Joseph},
	doi = {10.1103/PhysRevSTPER.9.020102},
	issue = {2},
	journal = {Phys. Rev. ST Phys. Educ. Res.},
	month = {Jul},
	numpages = {12},
	pages = {020102},
	publisher = {American Physical Society},
	title = {Assessing class-wide consistency and randomness in responses to true or false questions administered online},
	url = {http://link.aps.org/doi/10.1103/PhysRevSTPER.9.020102},
	volume = {9},
	year = {2013}}

@article{chen2025grading,
  title={Grading explanations of problem-solving process and generating feedback using large language models at human-level accuracy},
  author={Chen, Zhongzhou and Wan, Tong},
  journal={Physical Review Physics Education Research},
  volume={21},
  number={1},
  pages={010126},
  year={2025},
  publisher={APS}
}

@article{laverty12b,
	author = {James T. Laverty and Wolfgang Bauer and Gerd Kortemeyer and Gary Westfall},
	journal = {Phys. Teach.},
	pages = {540-543},
	title = {Want to Reduce Guessing and Cheating While Making Students Happier? Give More Exams!},
	volume = {50},
	year = {2012}}

@misc{Introduction,
	note = {[Online; accessed 2024-09-11]},
	title = {Introduction \textbar{}  {LangChain}},
	url = {https://python.langchain.com/v0.2/docs/introduction/},
	howpublished = {https://python.langchain.com/v0.2/docs/introduction/},
}
\bibliographystyle{rsc} 
\end{document}